\documentclass[epsfig,12pt]{article}
\usepackage{epsfig}
\usepackage{graphicx}
\usepackage{array}
\usepackage{color}
\usepackage{bm}

\usepackage{amsmath}
\usepackage{amsfonts}
\usepackage{graphicx}
\usepackage{cancel}
\usepackage{fullpage}
\usepackage{appendix}
\usepackage{enumerate}

\newcommand{\Lagr}{\mathcal{L}}

\newcommand{\der}[2]{\frac{\partial#1}{\partial#2}}

\newcommand{\non}{\nonumber\\}

\newcommand{\beq}{\begin{equation}}   
\newcommand{\eeq}{\end{equation}}
\newcommand{\beqn}{\begin{eqnarray}}   
\newcommand{\eeqn}{\end{eqnarray}}

\newcommand{\U}{\mbox{U}}
\newcommand{\SU}{\mbox{SU}}
\newcommand{\gsim}{\lower.7ex\hbox{$
\;\stackrel{\textstyle>}{\sim}\;$}}
\newcommand{\lsim}{\lower.7ex\hbox{$
\;\stackrel{\textstyle<}{\sim}\;$}}
\begin{document}

\begin{titlepage}

\begin{flushright}
FTPI-MINN-18/16, UMN-TH-3727/18\\
1/31/19

\end{flushright}

\vspace{5mm}

\begin{center}
{  \Large \bf  
Interpolating Between CP(\boldmath{$N-1$}) and \boldmath{$S^{2N-1}$} Target \\[2mm]
Spaces
}

\vspace{7mm}

{\large \bf   Daniel Schubring$^{\,a}$ and Mikhail~Shifman$^{\,b}$  }
\end {center}

\begin{center}
$^a${\it  Physics Department,
University of Minnesota,
Minneapolis, MN 55455}\\

$^b${\it  William I. Fine Theoretical Physics Institute,
University of Minnesota,
Minneapolis, MN 55455}

\end{center}

\vspace{10mm}

\begin{center}
{\large\bf Abstract}
\end{center}

Some magnetic phenomena in correlated electron systems were recently shown to be described in the continuum limit by 
a class of sigma models which present a U(1) Hopf fibration over CP(1). In this paper 
we study a generalization of such models with a target space given by a U(1) fibration over Grassmannian manifolds, of which CP($N-1$) is a special case. The metric of our target space is shown to be left-symmetric which implies that it is fully parametrized by two constants: the first one -- the conventional coupling constant --
is responsible for the overall scale while the second constant $\kappa$ parametrizes the strength of a deformation.
In two dimensions these sigma models are perturbatively renormalizable. We calculate their $\beta$ functions to two loops and find the RG flow of the coupling constants. We calculate the two-point function in the UV limit, which has a power law dependence with an exponent dependent on the RG trajectory.

\end{titlepage}

\newpage

\section{Introduction}
\label{intro}

Sigma models are used in theoretical physics as effective descriptions of a large number of phenomena -- from hadronic physics to condensed matter, to string theory. Probably the first physical application  
dates back to the 1960 work of Gell-Mann and L\'evy \cite{GML}. Since then, various aspects of the sigma models, including geometric,
have been thoroughly studied. 

Recently, it was rediscovered \cite{1one} that some noncollinear magnetic phenomena in correlated electron systems in the continuum limit are described by a sigma model on a target space with a geometry that interpolates between the two-dimensional sphere $S^2$ and the three-dimensional sphere $S^3$. The sigma model on $S^2$ is known variously as the $O(3)$, $CP(1)$, or classical Heisenberg model. The sigma model on $S^3$ likewise is known as either the $O(4)$ model or the $\SU(2)\times\SU(2)$ Principal Chiral Model (PCM). These two sigma models are known to be integrable in two spacetime dimensions and were exactly solved \cite{2two}-\cite{W4}.

There is a connection between $S^3$, thought of as the Lie group $SU(2)$, and $S^2$ through the well known Hopf fibration. Modding out a $U(1)$ subgroup of $SU(2)$ we recover $S^2$. If we incompletely mod out the $U(1)$ subgroup, giving the $S^2$ base space fibers of a small but nonzero size, we recover an interpolating geometry which may be called $SU(2)\times U(1)/U(1)$. In this paper we will generalize $S^2$ to arbitrary Grassmannian manifolds, and consider the target space $SU(N)\times U(1)/SU(M)\times SU(N-M)\times U(1)$.

If we pull back the metric of this target space to the Lie group $\SU(N)$ we find that the metric is left-invariant. The requirement of left-invariance restricts the number of parameters in the metric to just two. One is a parameter $\lambda$ characterizing the overall scale of the geometry. Such a parameter also appears in the $O(N)$, $CP(N-1)$, and PCM models, which are Einstein manifolds. The other parameter is the interpolation parameter $\kappa$, which measures the size of the $U(1)$ fibers. These two parameters can be viewed as coupling constants, since they characterize non-linearity of the model under consideration. All covariant characteristics, such as the Riemann and Ricci tensors can be expressed in these parameters.

In the process of submitting an early version of this paper, we discovered that sigma models on $SU(2)\times U(1)/U(1)$ and closely related spaces have been studied much earlier in the context of frustrated spin systems. See for example the review \cite{frustratedReview}, and some examples of early papers \cite{DombreRead}-\cite{ChubukovEtAl}. In particular, a 1995 paper by Azaria, Lecheminant, and Mouhanna \cite{MouhannaEtAl1995} has significant overlap with this paper. They also consider $U(1)$ fibered $CP(N-1)$ models, and they examine the model in $2+\epsilon$ spacetime dimensions, and in the large $N$ limit, which we will not discuss here. This paper differs in that we extend to arbitrary Grassmannian base spaces, and also in our focus on combined left invariance and gauge invariance as a principle restricting the space of parameters, and the discussion of multiple methods of finding the one-loop RG equations, each of which offers some advantage.

As in \cite{MouhannaEtAl1995}, the primary method used here to find the RG equations involves a short calculation based on the structure coefficients of the group $SU(N)$. We give a self-contained presentation of this method which we adapted from a paper by Milnor \cite{Milnor}. As a check of this method we also use an explicit coordinate system on the $U(1)$ fibered $CP(N-1)$ model in Appendix \ref{appendix coordinates}, and the connection coefficients found here may be of use in studying this geometry in other contexts. We also give a calculation directly in terms of loop integrals in the background field method. This method is naturally extended to find the two-point correlation function at one loop, and we note power law behavior in the UV which is quite distinct from that appearing in the limits of the $O(2N)$ and $CP(N-1)$ sigma models themselves.

\subsection{Basic construction of the model}

Here we will give a short introduction to the Lagrangian of the fibered $CP(1)$ model, showing how it reduces to the PCM and ordinary $CP(1)$ models in the appropriate limits. This Lagrangian will be discussed again from a slightly different point of view in Section \ref{FiberedCP} where it will be extended to all $N$.

The original motivation for this work was provided by \cite{1one} where it was noted that certain magnetic phenomena e.g. on the pyrochlore lattice in the continuum limit can be summarized by the
model
\beq
{\cal H} =\frac{1}{2\lambda^2}  \int d^D x \left\{ \left[\sum_{a=1,2,3 }  J_\mu^a J_\mu^a \right] - \kappa J_\mu^3 J_\mu^3\right\}
\label{cl1}
\eeq
where the current $J_\mu$ is defined as
\beq
J_\mu = -i U^\dagger \partial_\mu U \equiv \sum_{a} 2\, J_\mu^a T^a\,,\qquad
J_\mu^a = {\rm Tr}\, \Big( J_\mu  T^a\Big)\,.
\label{cl2}
\eeq
Here $U$ is an arbitrary $x$-dependent matrix, $U(x) \in \mbox{SU}(2)$, the generators are proportional to the Pauli matrices, $T^a=\tau^a/2$,  and $\kappa$ is a numerical parameter,
\beq
0 \leq \kappa \leq 1\,.
\label{cl3}
\eeq
If $\kappa =0$ this Lagrangian is just that of the $SU(2)$ PCM. But when $\kappa =1$ the term associated to the $J^3$ direction is canceled and the Lagrangian becomes that of the $CP(1)$ model.\footnote{Note that in another common convention for the CP(1) model, $2/g^2$ is the factor multiplying the Lagrangian. In this notation $\lambda^2=g^2/4$.}

To see why this is indeed the $CP(1)$ model, let us start from a particularly useful  formulation (the so-called gauged, or Witten,  formulation \cite{wit}) of the CP$(N-1)$ model. For the moment we will generalize to all $N$. The corresponding   Lagrangian
can be written as
\beq
{\cal L} =\frac{1}{2\lambda^2} \Big[D_\mu\bar{n} D_\mu {n} \Big]\,,\qquad D_\mu = \partial_\mu -iA_\mu\,, 
\label{cl4}
\eeq
where $n$ is an $N$-component complex scalar field $n^i$ ($i=1,2,..., N$) in the fundamental representation of the $\SU(N)$ group 
subject to the constraint
\beq
\bar{n} n =1\,.
\label{cl5}
\eeq
Moreover, $\lambda^2$ is a  constant.  Depending on the spacetime dimension $D=2,3,4$ it can have dimension of $[m^0],\,[m^{-1}]$, and $[m^{-2}]$.  Note, that (\ref{cl4}) has no kinetic term for the $A_\mu$ field. Eliminating $A_\mu$ by virtue of the equation of motion we arrive at
\beq
{\cal L} = \frac{1}{2\lambda^2} \Big[\partial_\mu\bar{n} \partial_\mu {n}  + (\bar{n}\partial_\mu n)^2\Big]. 
\label{cl6}
\eeq
Both Lagrangians (\ref{cl4}) and (\ref{cl6}) are U(1) gauge invariant. This is the reason why they describe CP$(N)= \SU(N)/\SU(N-1)\times\U(1)$  sigma model.

There is a rather obvious generalization of (\ref{cl4}), a ``mass'' term for $A_\mu$, which preserves the global symmetry of the model, namely
\beq
{\cal L} \to {\cal L}_\kappa = \frac{1}{2\lambda^2}\, \Big[D_\mu\bar{n} D_\mu {n} +\frac{1-\kappa}{\kappa} A_\mu^2\Big]\,. 
\label{cl7}
\eeq
Here $\kappa$ is a dimensionless parameter from the interval 
(\ref{cl3}). Now,
\beq
A_\mu= -i\frac \kappa 2 \left(\bar n \stackrel{\leftrightarrow}{\partial_\mu} n\right)
\label{mon1}
\eeq
and, therefore, 
\beq
{\cal L}_\kappa = \frac{1}{2\lambda^2}\, \Big[\partial_\mu\bar{n} \partial_\mu {n}  +  \kappa \left(\bar{n}\partial_\mu n\right)^2\Big]. 
\label{lagrangianN}
\eeq
If $\kappa =1$ we return to (\ref{cl6}). If $\kappa \neq 1$ the U(1) gauge symmetry is obviously lost.

This is the form of the fibered $CP(N-1)$ model in which the connection to the ordinary $CP(N-1)$ and also the $O(2N)$ sigma model is evident. To go back to the form \eqref{cl1} which is more natural for describing a PCM, let us choose a ``reference'' field configuration $n_0$,
\beq
n_0^i = 0 \,\, \mbox{for}\,\,  i=1,2,..., N-1 \,\, \mbox{and} \,\, n^{N} = 1\,.
\label{cl8}
\eeq
Then, in the most general case one can write
\beq
n(x) = U(x) n_0\,,\quad U\in \SU(N)\,, \quad N= 2,3, ...
\label{cl9}
\eeq
implying that
\begin{align}
{\cal L}_\kappa &=\frac{1}{2\lambda^2}\, \Big[\bar{n}_0\big(\partial_\mu U^\dagger  \partial_\mu U\big){n_0}  + \kappa \big(\bar{n}_0 U^\dagger \partial_\mu U n_0 \big) \big( \bar{n}_0 U^\dagger \partial_\mu U n_0 \big)\Big]\non
&=\frac{1}{2\lambda^2}\, \Big[-\bar{n}_0\left(J_\mu J_\mu\right)n_0  + \kappa \left(\bar{n}_0J_\mu n_0\right)\,\left(\bar{n}_0J_\mu n_0\right)\Big]
\label{mon2}
\end{align}
where the anti-Hermitian matrix $J_\mu$ is defined as in \eqref{cl2}. For low $N$ it is simple enough to use explicit formulas for the generators and structure coefficients to reduce this further. The final result for $N=2$ is precisely what is shown in (\ref{cl1}).

Extending this to the next most complicated case, $N=3$, we obtain
\beq
{\cal L}_\kappa=\frac{1}{2\lambda^2}\, \left(\,  \sum_{a=4}^{7} \, J_\mu^a  J_\mu^a   
+\frac{4}{3} (1-\kappa) J_\mu^8  J_\mu^8\,
\right)
\label{miri57}
\eeq
where the indices follow the standard convention of Gell-Mann matrices.
The model (\ref{miri57}) presents a continuous interpolation of the four-dimensional target space CP(2) to the five dimensional sphere $S^{5}$ through intermediate 
``squashed" $S^{5}$ at $\kappa<1$. In what follows in the general case we will denote these  spaces as $S_\kappa^{2N-1}$. Needless to say that topologically
$S_\kappa^{2N-1}$ is equivalent to $S^{2N-1}$.
Note that CP(2) is K\"ahlerian while $S^{5}$ is not.

We could of course go on to find the Lagrangian for general $N$ starting from \eqref{mon2}, but we will present this in a slightly different way in Section \ref{FiberedCP}. Already for the case $N=3$, notice that the currents for $a=1,2,3$ do not actually appear in the Lagrangian. This is implying a kind of gauge invariance which will be relevant to the case of general $N$. 

\subsection{Outline}

The organization of the paper is as follows.
 In Sect. \ref{FiberedCP} we discuss the fibered $CP(N-1)$ model from a more general point of view. In Sect. \ref{ind} the transition from the $n$ representation in \eqref{lagrangianN} to the $J$ representation given for example by \eqref{cl1} is discussed for general $N$, introducing the relevant concepts of left-invariance and gauge invariance along the way. In Sect. \ref{olr} we
will derive the one-loop renormalization equations for the model. This will be done via an explicit one-loop calculation. The advantage of this calculation is that we can easily find the anomalous dimension of the field $n$, which will allow us to find a new expression for the 2-point correlation function.

In Sect. \ref{FiberedGrass}, we extend the model to a base space which is a general Grassmannian. In Sect. \ref{FiberedGrassRenorm} the renormalization equations for this model are found up to two loops using a different approach. This involves finding the Ricci tensor for the fibered Grassmannian, which we will do using a method which takes advantage of the left-invariance property and the structure coefficients of the group $SU(N)$ rather than a direct approach using coordinates.

The mathematical basis for this method is outlined in Appendix \ref{Geometry}. $SU(N)$ can be described as fiber bundle over a base space which is the fibered Grassmannian (which is itself a fiber bundle), and the theory behind using the geometry of a fiber bundle to calculate properties of the base space is discussed in \ref{mfb}. The concrete formulas used to calculate the curvature in terms of properties of the Lie group are derived in Sec. \ref{crt}. Then, as a supplement, to illustrate the mathematical content of Appendix \ref{Geometry} in a more concrete setting, in Sec. \ref{appendix so3} we consider the group $SO(3)$ acting on $S^2$, and use the structure coefficients of the group to calculate the Christoffel connection coefficients and scalar curvature. 

The body of the paper finds the one-loop renormalization equations in two ways. A diagrammatic way in Sec. \ref{FiberedCP}, and an algebraic way in Sec. \ref{FiberedGrass} and the accompanying Appendix \ref{Geometry}. There is a third, geometric way which involves putting explicit coordinates on the fibered spaces we are interested in. Two choices of coordinates are discussed in Appendix \ref{Coordinate}.

Appendix \ref{appendix A} is devoted to the special case of the fibered $CP(1)$ model, in which we can use coordinates which are a natural extension of common coordinates on the PCM. In Appendix \ref{appendix coordinates} we introduce a natural extension of Fubini-Study coordinates which can be used for general $N$, and find the connection coefficients and Ricci tensor for the fibered $CP(N-1)$ model in these coordinates.

\section{Fibered $CP(N-1)$}
\label{FiberedCP}

The family of metrics we are considering is defined on the $2N-1$ dimensional unit sphere. As was mentioned we parametrize this with $N$ complex coordinates $n^i$, which are constrained to have unit norm (\ref{cl5}).
We will suppress the $i$ indices when there is no danger of confusion.
	
	The metric is defined implicitly through the Lagrangian (\ref{lagrangianN}) of a sigma model with parameters $\lambda, \kappa$.
		When  $\kappa=1$ the model becomes gauge invariant under transformations $n^i\rightarrow e^{i\phi(x)}n^i$, and it reduces to the sigma model on the complex projective space ${\rm CP}(N-1)$.
	
	For intermediate $\kappa$ we have a sigma model on a space with a less familiar metric. We can find the metric explicitly by transforming from $n^i$ to some unconstrained real coordinates $\phi^i$, in which case the Lagrangian becomes
	\begin{align}\Lagr = \frac{1}{2}g_{ij}(\phi)\partial_\mu \phi^i\partial^\mu \phi^j,\label{lagrangianPhi}\end{align}
	and we can read off the components of the metric $g_{ij}$ straightforwardly. This is the approach we will take in Appendix \ref{appendix coordinates}. But for now we will take a more abstract approach which might nevertheless illuminate why only two parameters are sufficient for this model.
		
	\subsection{Lifting to \boldmath${\SU(N)}$}
	\label{ind}
	
\subsubsection{Tangent vector \boldmath{$J$}}	
	Rather than considering the sigma model to live on the topological unit sphere, we will lift it to the Lie group $\SU(N)$ which acts on the unit sphere. Given a reference unit vector $n_0$, for each unit vector $n$ we can pick an element $U\in \SU(N)$ which transforms $n_0$ to $n$,
	\begin{align}
U n_0 = n.\label{Udef}
	\end{align}

The choice of $U$ for a given $n$ is clearly not unique. The subgroup of elements $V$ such that $Vn_0= n_0$ is isomorphic to ${\rm SU}(N-1)$. And for any $U$ satisfying \eqref{Udef} the element $UV$ also transforms $n_0$ to $n$. This construction is one way of realizing $\SU(N)$ as a fiber bundle over the base space $S^{2N-1}$ with fiber ${\rm SU}(N-1)$. We will occasionally refer to this group ${\rm SU}(N-1)$ as the \emph{vertical} subgroup or subalgebra depending on context.

In the context of the sigma model, $n(x)$ is spacetime dependent field and thus so is $U(x)$. As the individual coordinates $x^\mu$ are varied $U(x)$ traces out paths in $SU(N)$, and the tangent vectors to these paths should appear in the sigma model Lagrangian. So let us consider what form the tangent vectors will take. Nearby the point $x_0$, the path $U(x)$ can be expressed by
\begin{align}
U(x)=U(x_0)\exp\left[i (x-x_0)^\mu \tau_\mu(x) \right],
\end{align}for some set of Hermitian traceless matrices $\tau_\mu(x)$ which depend on $x$. The tangent vector at $x_0$ as $x^\mu$ is varied is just the left-invariant vector field in the Lie algebra associated to $\tau_\mu(x_0)$,
$$\tau_\mu(x_0)=-iU^\dagger(x_0) \partial_\mu U(x_0).$$
If we extend this formula to all $x$ (not just $x_0$) it is just the definition of the current $J_\mu$ used earlier \eqref{cl2}. The point here is that for each value of $\mu$, $J_\mu$ specifies the tangent vector in the target space $SU(N)$ as $x^\mu$ is varied. Thus expressing the Lagrangian in terms of $J$ should tell us something about the metric in a basis of left-invariant vector fields. Denoting a standard basis of left invariant vector fields as $\tau_a$, we can find the components $J^a_\mu(x_0)$ of the tangent vector in this basis,

\begin{align}J^a \tau_a=-iU^\dagger\partial U.\label{Jdef}\end{align} 	
	Here we are suppressing spacetime indices and coordinates, and blurring the distinction between left invariant vector fields and the Hermitian traceless matrices with which they are associated.

	\subsubsection{Lagrangian in terms of \boldmath{$J$}}
	\label{lit}
	
	In the following we will choose the reference unit vector $n_0$ in \eqref{Udef} to be nonzero only in the last component. \begin{align}n_0^{N}=1, \text{and } n_0^i=0 \text{ otherwise.}\end{align}
	
Considering $U$ concretely as a matrix, the condition \eqref{Udef} fixes the last column of $U$ to be the vector $n$. The other columns may be freely chosen up to the constraint that $U$ be a unitary matrix. The other $N-1$ column vectors in $U$ are denoted by $e_{(i)}$,
	\begin{align}
	U=	\left(\begin{array}{ccccc}
	e_{(1)} & e_{(2)} &\dots & e_{(N-1)} & n
	\end{array}\right).
	\end{align}			
Then \eqref{Jdef} gives an expression for the components of the matrix $J$ in terms of $n$ and $e_{(i)}$,\begin{align}
	J_{ij} = -i\left(\begin{array}{cc}
	e_{(i)}^{\dagger}\partial e_{(j)} &e_{(i)}^{\dagger}\partial n \\[2mm]
	n^\dagger\partial e_{(j)} & n^\dagger\partial n
	\end{array}\right).\label{JExplicit}
	\end{align}	
	Since this must be traceless we have the identity
	\begin{align}
	\sum_i 	e_{(i)}^{\dagger}\partial e_{(i)}=- n^\dagger\partial n.\label{traceIdentity}
	\end{align}
	
	Let us now rewrite the Lagrangian \eqref{lagrangianN} in terms of components of $J$, which again describe the motion in the $\SU(N)$ target space rather than $S^{2N-1}$.
	
Note that the columns of a unitary matrix are orthonormal. Hence,  
	\begin{align}
	e_{(i)}^{\dagger}e_{(j)}=\delta_{ij},\qquad e_{(i)}^{\dagger}n=0\,.
	\end{align}
	This means that $n,\, in,\, e_{(i)},\, ie_{(i)}$ form a complete orthonormal basis of $C^N$ considered as a real vector space with metric $\langle z,w\rangle\equiv\text{Re}(z^\dagger w)$, which is the ordinary Euclidean metric if we identify this space with $R^{2N}$. So we can expand $\partial n$ in terms of this complete basis.
\begin{align}
\partial{n}&=\text{Re}(-in^\dagger \partial n)in+\text{Re}(e_{(i)}^{\dagger} \partial n)e_{(i)}+\text{Re}(-ie_{(i)}^{\dagger} \partial n)ie_{(i)}\non
&=(n^\dagger\partial n)n+(e_{(i)}^{\dagger} \partial n)e_{(i)}\, ; \\
|\partial n|^2 &= |n^\dagger\partial n|^2+\sum_i |e_{(i)}^{\dagger} \partial n|^2\,.\label{dn2Components}
\end{align}
Thus the Lagrangian \eqref{lagrangianN} becomes,
\begin{align}
\Lagr&= \frac{1}{2\lambda^2}\left[(1-\kappa)|n^\dagger\partial n|^2+\sum_i |e_{(i)}^{\dagger} \partial n|^2\right].\label{lagrangianE}
\end{align}	

\subsubsection{Lie algebra basis}
\label{lab}

Now the Lagrangian is written in terms of components of $J$ in \eqref{JExplicit}, but to proceed, let us choose a standard basis on the Lie algebra. For convenience notating the dimension of the Lie subgroup $SU(N-1)$ as $M$,\begin{align}M \equiv (N-1)^2-1,\end{align} the first $M$ Lie algebra elements $\tau_a$ belong to the vertical subalgebra that keeps $n_0$ invariant. As matrices, both the $N$-th row and column vanish.

The next $2(N-1)$ Lie algebra elements vanish everywhere except on the $N$-th row and column. Moreover, $\tau_{M+2k-1}$ has a form similar to the Pauli matrix 
$\sigma^1$, with a 1 in the $k$-th position of the last row and column, and $\tau_{M+2k}$ has a form similar to $\sigma^2$ with an $i$ and $-i$ in those positions respectively,
\begin{align}
(\tau_{M+2k-1})_{ij}&=\delta_{iN}\delta_{jk}+\delta_{ik}\delta_{jN}\,,\non [2mm]
(\tau_{M+2k})_{ij}&=i\delta_{iN}\delta_{jk}-i\delta_{ik}\delta_{jN}\,.\label{tauLastRowColumn}
\end{align}
Finally the last Lie algebra element is diagonal and commutes with the ${\rm SU}(N-1)$ subalgebra,
\begin{align}
\tau_{N^2-1}=\sqrt{\frac{2}{N(N-1)}}\text{diag}\left(1, 1, \dots, 1, -(N-1)\right).
\end{align}
This standard basis is chosen so that the structure coefficients are completely antisymmetric, and so that the basis matrices satisfy the trace identity
\begin{align}
\text{Tr}(\tau_a\tau_b)=2\delta_{ab}.\label{traceOrthogonal}
\end{align}

For the sake of discussing these Lie algebra elements, we will refer to the first $M$ elements in the $\SU(N-1)$ subalgebra as \emph{vertical} elements. The remaining directions are referred to as \emph{horizontal}. The horizontal elements may be further distinguished between those of the form \eqref{tauLastRowColumn} which we refer to as \emph{K\"ahler} elements, and $\tau_{N^2-1}$ which we refer to as the \emph{phase} element. As we shall soon see, when $\kappa=1$ and the model becomes CP$(N-1)$ only these so-called  K\"ahler elements will appear in the Lagrangian.

\subsubsection{Left invariance}
\label{lin}

Now we can find the components of $J$ in this basis by using the explicit form for $J$ in \eqref{JExplicit}, and taking traces using \eqref{traceOrthogonal},
\begin{align}
J^{M+2k-1}&=\text{Im}(e_{(k)}^\dagger\partial n)\,,\\[2mm]
J^{M+2k}&=\text{Re}(e_{(k)}^\dagger\partial n)\,,\\[2mm]
J^{N^2-1}&=\sqrt{\frac{N}{2(N-1)}}in^\dagger\partial n\,,\label{Jlastcomponent}
\end{align}
where in \eqref{Jlastcomponent}, the identity \eqref{traceIdentity} was used.

Our Lagrangian \eqref{lagrangianE} now becomes quite simple in this basis,
\begin{align}
\Lagr&= \frac{1}{2\lambda^2}\left[\sum^{2(N-1)}_{m=1} (J^{M+m})^2+(1-\kappa)\frac{2(N-1)}{N}(J^{N^2-1})^2\right].\label{lagrangianJ}
\end{align}
As in \eqref{lagrangianPhi}, this sigma model Lagrangian is just the metric on the target space contracted with the tangent vector to the path traced out by the field. So in this left-invariant basis, the metric is diagonal and does not depend on position on the target space. This means that the class of metrics we are considering itself has the property of \emph{left invariance}. If we know the metric at one point on the target space, we can use left translation to pull back the metric to any other point. In particular this means the space is homogeneous, and the Ricci scalar should not depend on position.

Note that since there is no appearance of the components in the vertical directions this metric is degenerate, i.e. it vanishes acting on the vectors in the vertical directions. This will lead to problems in naively applying results from Riemannian geometry.

\subsubsection{Gauge invariance}
\label{gin}

Considered as a metric on $\SU(N)$ there is one other important property this metric has, and that is what we will call \emph{gauge invariance} in this context. As mentioned previously, our field $U(x)$ in $\SU(N)$ is not unique, and we can multiply on the right by an arbitrary space dependent member of the subgroup $V(x)\in \SU(N-1)$. An equivalent way to consider this is that we are allowed to arbitrarily choose a distinct orthonormal basis $e_{(i)}(x)$ at each spacetime point, and this choice will change the components $J^a$ that appear in our Lagrangian. 

If we transform $U\rightarrow UV$, our $J=-iU^\dagger\partial U$ vector transforms to,
\begin{align}
J\rightarrow V^\dagger J V-iV^\dagger\partial V.
\label{miri55}
\end{align}
The inhomogeneous term is a member of the vertical subalgebra. It will arbitrarily change the vertical components $J^{a\leq M}$. If the Lagrangian is not to depend on choice of $V,$ these components must not appear in the Lagrangian. In other words, if a left-invariant metric is to be gauge invariant, it must be degenerate and vanish when acting on vectors from the subalgebra.

So the only allowed terms in the Lagrangian are quadratic in the remaining components $J^{a>M}$ and they must be invariant under the adjoint transformation $V^\dagger J V$. There are only two independent terms which satisfy this. Since $\tau_{N^2-1}$ commutes with the subalgebra, $J^{N^2-1}$ is a scalar under gauge transformations, and $$\left(J^{N^2-1}\right)^2$$ is one allowed term. This can also be seen from \eqref{Jlastcomponent}, where there is no dependence on $e_{(i)}$. 

If we complexify the Lie algebra, the  K\"ahler elements break up into two fundamental representations with basis elements $\tau_{M+2k-1}\mp i\tau_{M+2k}$, which correspond to those matrices which are nonzero only in the last row and column respectively (similar to raising and lowering matrices in $SU(2)$). The quadratic invariant in both these representations is just the sum over the  K\"ahler directions $$\sum^{2(N-1)}_{m=1}\left(J^{M+m}\right)^2,$$see Eq. (\ref{miri57}) as an example for $SU(3)$.

So there are only two independent terms possible for a sigma model Lagrangian satisfying left invariance and gauge invariance, and thus as long as these properties are preserved under renormalization we only need two parameters, $\lambda$ and $\kappa$. We will show this explicitly to one loop.
	
\subsection{One-Loop Renormalization (\boldmath{$D=2$})}
\label{olr}

Nothing so far has depended on the spacetime dimension, as we have been primarily focused on the geometry of the target space. But now we will specialize to two
dimensions, in which it is well known that the one-loop renormalization of the sigma model is given by the Ricci flow (see e.g. \cite{HonerkampEcker},\cite{Ketov}). If $\mu$ is the scale at which we define our parameters in the metric $g$, and $R$ is the Ricci tensor,
\begin{align}
\mu\der{}{\mu}g_{\alpha\beta}(\kappa,\lambda)=\frac{1}{2\pi}R_{\alpha\beta}(\kappa,\lambda).
\label{renormGroup}
\end{align}
So solving the problem of finding the renormalization to one loop amounts to the purely geometrical task of finding the Ricci tensor.

One straightforward way of approaching this is to introduce coordinates on the target space, which allows us to find the components of the metric via the general form of the sigma model Lagrangian \eqref{lagrangianPhi}. If we can invert this metric, we can calculate connection coefficients and the Ricci tensor by a tedious but straightforward calculation.

One possible set of coordinates involves an overall phase $\phi$ given by the first component on the unit sphere 
\beq
n^0=|n^0|e^{i\phi}\,,
\eeq
and the remaining real coordinates $x^i, y^i$ are given by the real and imaginary parts of the Fubini-Study coordinates on ${\rm CP}(N-1)$,
\beq
x^i+i y^i=\frac{n^i}{n^0}\,.
\label{miri58}
\eeq

This coordinate system has the advantage that the coordinate vector $\partial_\phi$ is directly related to what we are calling the phase element, $\tau_{N^2-1}$,
\begin{align}
\partial_\phi=-\sqrt{\frac{N}{2(N-1)}}\tau_{N^2-1},\label{phiCoord}
\end{align}
and when $\kappa=1$ all expressions reduce to those in the well-known Fubini-Study coordinates. Expressions for the metric, inverse metric, and connection coefficients in this coordinate system are given in Appendix \ref{appendix coordinates}.

Rather than the straightforward but tedious coordinate method, we will consider two other means of calculation in the body of the paper. Later on in the context of the Grassmannian we will present an algebraic method to find the Ricci tensor which takes advantage of the left-invariance property discussed earlier. But for now we will momentarily forget the general solution \eqref{renormGroup}, and directly calculate loops in a version of Wilsonian renormalization adapted from Polyakov \cite{PolyakovONmodel}. The advantage of this method is we can also easily use it to find how the field scales under renormalization.

The bare action is originally defined in terms of a complex unit vector field $n_0$, which is defined with a hard momentum cutoff at scale $M_{UV}$. In order to consider the action in terms of the field $n$ defined with a lower cutoff $\mu$, we decompose $n_0$ in terms of $n$ and the orthonormal basis $e_{(i)}$, which we considered earlier in section \ref{lit} as column vectors of $U(x)$.
\begin{align}n_0^{a}=e^{i\sigma}\sqrt{1-|\phi|^2}n^a+\phi^i e^a_{(i)}\label{nLambda}\end{align}
The real field $\sigma$ and the $N-1$ complex component fields $\phi^i$ will be the fields we integrate over to find the action in terms of the background field $n$.

As a sidenote, one might ask why we do not use what might appear to be a simpler renormalization scheme. Rather than dealing with a constrained field $n_0$ and a gauge-dependent basis $e_{(i)}$, instead one might use some set of unconstrained coordinates on the target space $\phi_0$ as in \eqref{lagrangianPhi}, and decompose this linearly into a background field $\phi_b$ and a field we integrate over $\phi_q$,
$$\phi_0=\phi_b+\phi_q.$$
If the procedure is valid, one can even choose $\phi_b$ to take a special form to simplify the calculation. This method has for instance been shown in detail to work for the $O(3)$ model \cite{SigmaModelsQCD1984}. But unfortunately this scheme maintains neither manifest $SU(N)$ invariance, nor manifest covariance under diffeomorphisms of the target space, and it does not give the correct result for any $O(N)$ model with $N\neq 3$, at least without further modification. Curiously though, this $\phi_0$ background-field method can be shown to be valid for any K\"ahler target space manifold. 

Returning to the Polyakov-style scheme \eqref{nLambda}, we can express the original Lagrangian \eqref{lagrangianN} in terms of $n_0$ in terms of $\sigma,\phi^i$ and the background fields $n,e_{(i)}$. In doing so we will encounter elements of the matrix $J_{ij}$ as in \eqref{JExplicit}. We will give these elements names to emphasize the similarity to Polyakov's notation,
	\begin{align}
	A_{ji} \equiv e^\dagger_{(j)}\partial e_{(i)},\qquad B_i &\equiv e^\dagger_{(i)}\partial n,\qquad C\equiv n^\dagger\partial n.
	\end{align}
Now expanding the Lagrangian to second order in $\sigma,\phi^i$, and ignoring terms which will only lead to irrelevant terms at one loop, we find,
\begin{align}
	\Lagr=\frac{1}{2\lambda_0^2} &\{|B|^2 + (1-\kappa)|C|^2 + (1-\kappa)(\partial\sigma)^2+|\partial\phi|^2\label{kappaONterms0}\\
	&-\left(|B|^2+(1-2\kappa)|C|^2\right)|\phi|^2\label{kappaONtermsTadpole}\\
	&+(1-2\kappa)|\phi^i B_i^\dagger|^2+2(1-\kappa)i\partial\sigma\left(\phi^i B_i^\dagger+\phi^{i\dagger}B_i\right)\label{kappaONtermsSigmaPhi}\\
	&+\left(\frac{1}{(N-1)^2}+\frac{2\kappa}{N-1}\right)|C|^2|\phi|^2+C\left(\frac{1}{N-1}+\kappa\right)\left(\phi^\dagger\partial\phi-\partial\phi^\dagger\phi\right)\}.\label{kappaONtermsGauge}
	\end{align}
	The first two terms in the first line \eqref{kappaONterms0} take the form of the original Lagrangian \eqref{lagrangianE} in terms of the background fields, and the second two terms give the propagators for $\sigma$ and $\phi^i$. Now we can integrate out $\sigma, \phi^i$ at one loop, leading to the renormalized Lagrangian,
	\begin{align}
	\Lagr &=\frac{1}{2\lambda_0^2} \left(|B|^2 + (1-\kappa)|C|^2\right)-\frac{1}{2\pi}\log\frac{M_{UV}}{\mu}\left((N-1+\kappa)|B|^2+(N-1)(1-\kappa)^2|C|^2\right).
	\end{align}

\begin{figure}[t]
	\includegraphics[width=0.6\textwidth]{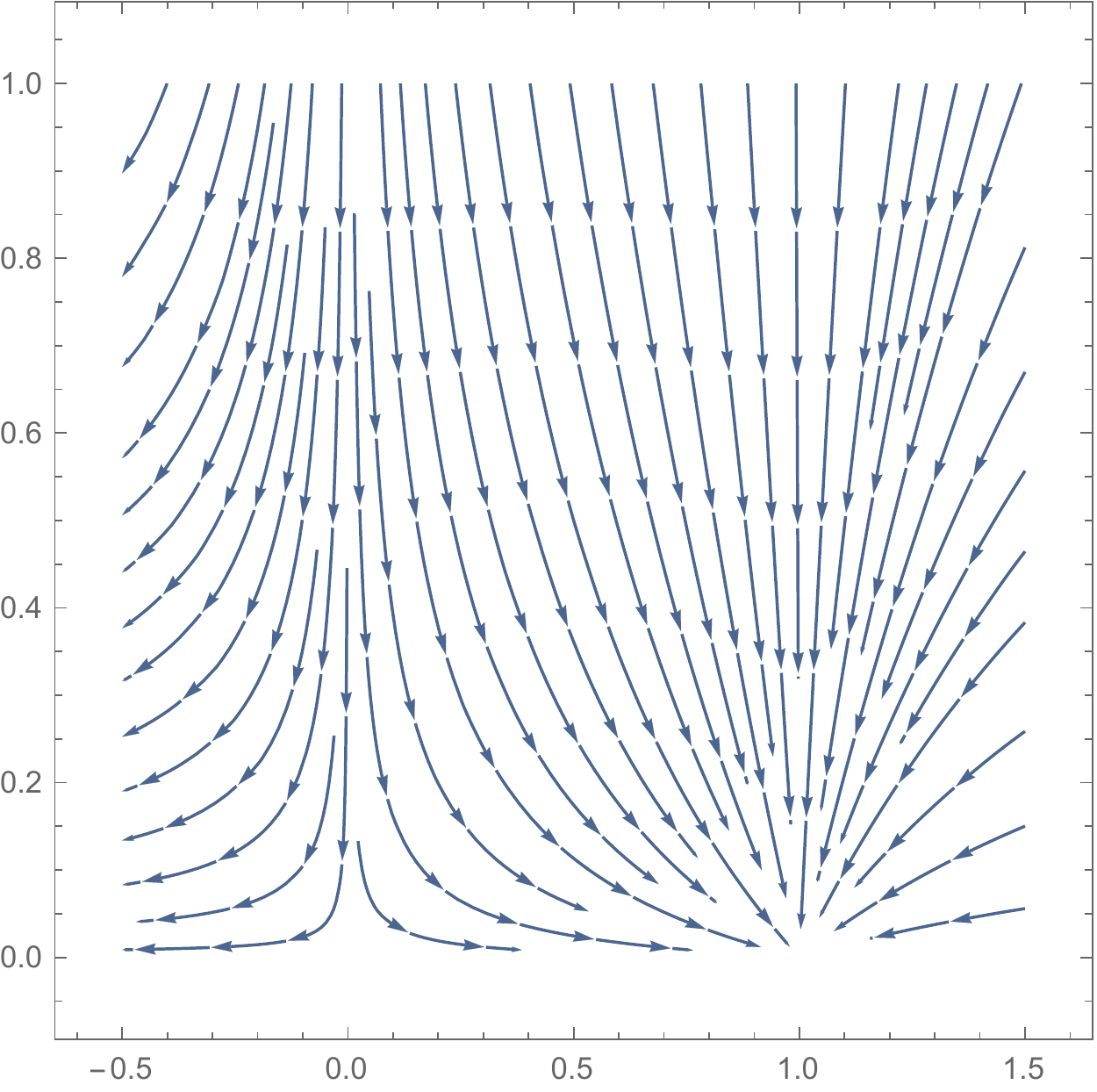}
	
	\small RG flow for the $S^{3}_\kappa$ model. Higher $N$ are qualitatively similar. The arrows are pointing towards the UV. The vertical axis is the ordinary coupling constant $\lambda^2$ multiplying the Lagrangian in both the SU(2) PCM ($\kappa=0$) and the $CP(1)$ sigma model ($\kappa=1$). The horizontal axis is the interpolation parameter $\kappa$. The physical region is between $0\leq \kappa\leq 1$.  Trajectories near the left side of the plot that pass near the PCM fixed point at (0,0) have small values of the parameter $K$. 
\end{figure}

We can easily read off the beta functions for $1/\lambda^2$ and $(1-\kappa)/\lambda^2$, which are identical to the one loop beta functions for the parameters $\eta_1$ and $\bar{\eta}_2$ appearing first in \cite{MouhannaEtAl1995}. Rewriting the beta functions in terms of the parameters $\lambda^2$ and $\kappa$,
\begin{align}\mu\der{}{\mu} {\lambda^2}&=-\frac{\lambda^4}{\pi}(N-1+\kappa),\label{lambdaFlow}
\\\mu\der{}{\mu} \kappa&=\frac{\lambda^2}{\pi}N\kappa(1-\kappa).\label{kappaFlow}
\end{align}	
In particular, we see that for $\kappa=0$ or $\kappa=1$, corresponding to the $O(2N)$ and CP$(N-1)$ sigma models respectively, the parameter $\kappa$ does not run. And the renormalization group equation for $\lambda$ reduces to the known result for these models. For $0<\kappa<1$, the parameters flow to the stable fixed point $\lambda=0, \kappa=1$ as the renormalization scale $\mu$ increases toward the UV.

As is usual for asymptotically free theories, the dimensionless bare parameter $\lambda^2$ will be replaced by a dimensionful parameter $\Lambda$ which sets the scale for the spectrum and correlation lengths. We did not find this independently and will not make use of this in the following, but note it was found in \cite{MouhannaEtAl1995}.

However there is another RG invariant parameter, also first found by \cite{MouhannaEtAl1995}. Note that we can divide the beta functions to find the slope of a RG trajectory $\lambda^2(\kappa)$, and then we can integrate to find a relation between $\lambda^2$ and $\kappa$ in terms of a new constant $K$ that parametrizes the RG trajectories,
\begin{align}
K=\frac{\kappa^{1-\frac{1}{N}}}{1-\kappa}\lambda^2.\label{Kdef}
\end{align} We will show that this parameter $K$ is essentially the anomalous dimension of the field $n$ about the UV fixed point.

Considering again the renormalization scheme \eqref{nLambda}, and integrating out the $\sigma, \phi^i$ fields to one loop in correlation functions involving $n_0$,
\begin{align}
\langle n_0\rangle&=\left\langle \left(1+i\sigma -\frac{1}{2}\sigma^2+\dots\right)\left(1-\frac{1}{2}|\phi|^2+\dots\right)n\right\rangle\non&=\left[1-\frac{\lambda^2}{4\pi}\left(\frac{1}{1-\kappa}+2(N-1)\right)\log \frac{M_{UV}}{\mu}\right]\langle n\rangle.\label{fieldRenorm}
\end{align}
Of course non-perturbatively $\langle n\rangle$ vanishes, but this field renormalization factor should also appear in correlation functions of multiple fields $n(x)$ at different spacetime points, as long as the distances are much larger than the cutoff scale. As usual this field renormalization can be used in the Callan-Symanzik equation along with the running couplings to find improved perturbation theory estimates for correlation functions.

In particular, due to dimensional analysis, the two-point function has the form
\begin{align}
\langle n^\dagger(p)\cdot n(-p)\rangle=\frac{1}{p^2}f\left(\frac{p^2}{\Lambda^2}\right)
\end{align}
where $f$ is some scaling function. Using \eqref{fieldRenorm}, we can write an RG equation for $f$ similarly to \cite{PolyakovONmodel},
\begin{align}
\frac{d\log f}{d\log (p/\mu)}=\frac{\lambda^2}{2\pi}\left(\frac{1}{1-\kappa}+2(N-1)\right)\rightarrow \frac{K}{2\pi}.
\end{align}
The limit here is taken at large momentum near the $\lambda^2=0,\,1-\kappa=0$ critical point. We can use \eqref{Kdef} to express the ratio $\lambda^2/(1-\kappa)$ in terms of the RG invariant $K$. So in the UV the correlation function shows power law behavior with an exponent that depends on the RG trajectory,
\begin{align}
\langle n^\dagger(p)\cdot n(-p)\rangle_{UV}\sim\frac{1}{p^2}\left(\frac{p^2}{\Lambda^2}\right)^\frac{K}{4\pi}.\label{powerLaw}
\end{align}

This behavior of the correlation function is distinct from both the $CP(N-1)$ model (in which this correlation function is not gauge invariant) and the $O(2N)$ model, and so it is something intrinsic to the interpolating model. As discussed in \cite{MouhannaEtAl1995}, for small $K$ the trajectory passes near the $O(2N)$ asymptotically free fixed point in the regime in which perturbation theory is still valid. For these trajectories there should be a regime in which the perturbative expression for the two-point function for the $O(2N)$ model \cite{PolyakovONmodel} is valid,
\begin{align}
\langle n^\dagger(p)\cdot n(-p)\rangle_{O(2N)}\sim\frac{1}{p^2}\left(\log\frac{p^2}{\Lambda^2}\right)^\frac{2N-1}{2N-2}.
\end{align}
This holds for $K$ small and $p$ much greater than $\Lambda$, but not large enough to leave the vicinity of the $O(2N)$ UV fixed point. For even larger $p$ there is a cross-over to the new power law behavior governed by the $CP(N-1)$ UV fixed point \eqref{powerLaw}. This new expression should hold for trajectories with large $K$ as well, in which case the perturbative expression for the $O(2N)$ model is not valid for any scale.

\section{Fibered Grassmannian}
\label{FiberedGrass}

Now we will generalize the fibration over complex projective space to a fibration over a general Grassmannian manifold. Complex projective space $CP(N-1)$ can be thought of as the space of all one-dimensional complex linear subspaces of $\mathcal{C}^N$. Similarly the Grassmannian $Gr(M,N)$ is the dual space of all $M$-dimensional linear subspaces of $\mathcal{C}^{N}$. As is well known, this is equivalent to the dual space $Gr(N-M,N)$. For convenience, the dimension of the dual manifolds is labeled $L\equiv N-M$,
$$L+M=N.$$
 The symmetry between $L$ and $M$ will eventually be manifest, but for the moment consider a representation of the $M$-dimensional linear subspaces in terms an orthonormal basis of $M$ linearly independent $N$-dimensional column vectors, $n^{i}_{(\alpha)}$. The Latin $i$ index runs from $1$ to $N$ and the Greek $\alpha$ index runs from $1$ to $M$. This can be thought of as a rectangular matrix with $M$ column vectors,
$$n=\left(n_{(0)}, \cdots, n_{(M)}\right).$$
Since a change of basis does not change the linear subspace, there should be an equivalence relation under multiplying $n$ on the right by a unitary matrix $V_M\in U(M)$. This will be manifested as a gauge symmetry in the Lagrangian, which is an extension of how the Lagrangian for the $CP(N-1)$ model involved $U(1)$ gauge symmetry.

The Lagrangian is defined in terms of the auxiliary $U(M)$ gauge field $A$,
\begin{align}\Lagr = \frac{1}{2\lambda^2}\text{Tr}_M\left[\left(\partial_\mu n^\dagger +i A_\mu n^\dagger \right)\left(\partial_\mu n -i n A_\mu\right)\right].\label{lagrGrass}\end{align}

	This can be treated similarly to the $CP(N-1)$ model by choosing a matrix $U\in SU(N)$ that maps a standard $n_0$ to $n$,
	$$U n_0 = n,\qquad n_0\equiv\left(\begin{array}{c}
0_{L\times M}\\
I_{M}
\end{array}\right).$$
Here $	I_{M}$ is the $M\times M$ identity matrix, and $0_{L\times M}$ are extra zeros to fill out the full $N\times M$ matrix.

This defining condition on $U$ fixes the last $M$ columns to be $n^i_{(\alpha)}$, but there is still a $SU(L)$ gauge freedom in picking the first $L$ columns $e^i_{(\alpha)}$.
$$U=\left(e_{(0)}, \dots, e_{(L)}, n_{(0)}, \dots, n_{(M)}\right).$$

Explicitly this is gauge symmetry under multiplying $U$ on the right by a matrix $V_L\in SU(L)$ that leaves $n_0$ invariant,
$$ U\left(\begin{array}{cc}
V_L& 0 \\
0 & I_M 
\end{array}\right)n_0= Un_0.$$

Besides the $SU(L)$ gauge invariance in the definition of $U$, there is also the original $U(M)$ gauge invariance appearing in the Lagrangian, which can be split into a $SU(M)$ part,
$$Un_0 V_M=U\left(\begin{array}{cc}
I_L& 0 \\
0 & V_M 
\end{array}\right)n_0\sim Un_0,$$
and a $U(1)$ part, which we again call the \emph{phase} part,
$$U\left(\begin{array}{cc}
\exp(-\frac{i}{L}\phi)I_L& 0 \\
0 & \exp(\frac{i}{M}\phi)I_M 
\end{array}\right)n_0\sim U n_0.$$

Under the substitution $n=Un_0$, the Lagrangian becomes,
	 \begin{align*}
	 \Lagr&=\frac{1}{2\lambda^2}\text{Tr}_M\left(n_0^\dagger(J_\mu-A_\mu)^2 n_0\right),\end{align*}
where as before $J_\mu=-i U^\dagger \partial_\mu U$. Integrating out the auxiliary gauge field removes the $SU(M)\times U(1)$ components of $J_\mu$,
 	\begin{align*}
 	\Lagr&=\frac{1}{2\lambda^2}\sum_{a,b\notin U(M)}J^{a\mu}J^b_\mu\text{Tr}_M\left(n_0^\dagger\tau_a\tau_b n_0\right),\end{align*}
 	and the appearance of $n_0$ removes the $SU(L)$ components. The only remaining components are the $2LM$ off-block-diagonal components of the Lie algebra, which we again call the \emph{K\"ahler} components.

	Thus the general Grassmannian Lagrangian can be written as,
\begin{align}
\Lagr&=\frac{1}{2\lambda^2}\sum_{a\in\text{K\"ahler}}\left(J_\mu^{a}\right)^2.\label{lagrGrassJ}\end{align}
In this form $L$ and $M$ are treated on a manifestly equal footing, and we could of course reverse the previous steps to express the Lagrangian in terms of the $N\times L$ matrix $e^i_{(\alpha)}$, instead of $n^i_{(\alpha)}$. 

As before, we will generalize this K\"ahler manifold by no longer gauging over the $U(1)$ phase subgroup. In the same way as for the $CP(N-1)$ model, it is easy to write the Lagrangian in terms of the only two left-invariant terms which are gauge invariant under $SU(L)\times SU(M)$,
\begin{align}\Lagr&=\frac{1}{2\lambda^2}\left(\sum_{a\in \text{K\"ahler}}(J^a)^2+\frac{2 (\mathcal{N}-1)}{\mathcal{N}}(1-\kappa)(J^\text{phase})^2\right),\label{lagrangianGrassJ}\end{align}
where by definition,
\begin{align}
\mathcal{N}&\equiv LM+1.
\end{align}
The unusual factor multiplying $J^{\text{phase}}$ is chosen for later convenience. Note that indeed $\mathcal{N}=N$ when $L=1$ and $M=N-1$, and so this normalization agrees with the fibered $CP(N-1)$ model \eqref{lagrangianJ} introduced earlier.

\subsection{One-Loop Renormalization}\label{FiberedGrassRenorm}

Now rather than doing explicit loop calculations, we will make use of the well-known one-loop expression in terms of the Ricci tensor \eqref{renormGroup}, and use a method of calculating the Ricci tensor which takes advantage of the left-invariance property.

The idea is rather than considering the metric $g$ on the Grassmannian itself, we consider the metric $\bar{g}$ pulled back to the Lie group $SU(N)$, which is left-invariant but degenerate in the vertical $SU(L)\times SU(M)$ subgroup directions. Concretely, the metric in the left-invariant basis $\tau_a$ is diagonal,
\begin{align}
\bar{g}(\tau_a,\tau_b)=\frac{1}{\lambda^2}C_a \delta_{ab},\label{metricLIBody}
\end{align}where from \eqref{lagrangianGrassJ} we have that $C_a=0$ for the vertical directions, $C_a=1$ for the  K\"ahler directions, and
\begin{align}
C_{\phi}=(1-\kappa)\frac{2(\mathcal{N}-1)}{\mathcal{N}},\label{metricPhaseBody}
\end{align}
for the phase direction.

The curvature of $\bar{g}$ can then be calculated borrowing an idea from Milnor \cite{Milnor}. The Lie bracket of the basis $\tau_a$ considered as left-invariant vector fields is directly related to the commutator of $\tau_a$ considered as matrices in the Lie algebra. This will ultimately allow us to determine the metric-compatible connection on the manifold in terms of the structure coefficients $f$ of the group, defined by,
\begin{align}
[\tau_a,\tau_b]=2i\sum_c f_{abc}\tau_c.\label{tauBracketBody}
\end{align}

  In the present case there are some subtleties in dealing with the degenerate directions of $\bar{g}$ and applying the results to the Grassmannian manifold we are interested in rather than $SU(N)$. But ultimately we will be able to express the components of the Riemann tensor only in terms of the diagonal components $C$ of the metric, and the structure coefficients of the group.
  
   This is not the first time something similar to this has been done. In particular a formula for the Riemann tensor in terms of structure coefficients was also found in \cite{AzariaEtAl1993} and used for the fibered $CP(N-1)$ model in \cite{MouhannaEtAl1995}. However both the derivation and particular formula used in this paper differs considerably from \cite{AzariaEtAl1993}. Here we focus on the fiber bundle structure induced for example by the map from $SU(N)\rightarrow Gr(M,N)$, and also the properties of the degenerate metric pulled back to $SU(N)$. The mathematical details are discussed in Appendix \ref{Geometry}, and here we will simply present the result. The diagonal components of the Ricci tensor (no summation is implied over $a$) are,
\begin{align}
R_{aa}	&=\sum_{b,c}f_{abc}^2\left(1+\frac{C_b-C_a}{C_{{c}}}\chi_c+3\frac{C_a-C_c}{C_{{b}}}\chi_b-\frac{C_b-C_a}{C_{{c}}}\frac{C_a-C_c}{C_{{b}}}\chi_b\chi_c\right),\label{mainFormulaBody}
\end{align}
where $\chi_a$ is just an indicator function that vanishes when $C_a=0$ and is $1$ otherwise.

Now simply inserting the structure coefficients for $SU(N)$ and the diagonal metric components \eqref{metricLIBody}\eqref{metricPhaseBody}, we find the Ricci tensor components for the K\"ahler directions,
\begin{align}
R_{aa}= 2(\mathcal{N}-1+\kappa)\frac{N}{\mathcal{N}},
\end{align}
and the phase direction,
\begin{align}
R_{\phi\phi}= N\left(\frac{2(\mathcal{N}-1)}{\mathcal{N}}\right)^2(1-\kappa)^2,
\end{align}
where again, $\mathcal{N}=LM+1$, which is convenient notation because $\mathcal{N}=N$ in the complex projective case.

Then using the expression for the beta function in terms of the Ricci tensor \eqref{renormGroup}, and again using the diagonal components of the metric \eqref{metricLIBody}, we find the one-loop RG equations for $\lambda, \kappa$.

\begin{align}
\mu\der{}{\mu}\lambda^2&=-\frac{\lambda^4}{\pi}(\mathcal{N}-1+\kappa)\frac{N}{\mathcal{N}},\\
\mu\der{}{\mu}\kappa&=\frac{\lambda^2}{\pi}\,N\kappa(1-\kappa).\label{RGKappaGrass}
\end{align}
Note that these RG equations of course reduce to those of the fibered $CP(N-1)$ model \eqref{lambdaFlow}\eqref{kappaFlow} when $\mathcal{N}=N$, and they reduce to that of the ordinary Grassmannian model when $\kappa=1$. When $\kappa=0$, the model reduces to a new Einstein manifold which is not equivalent to the $O(2N)$ model. 

\subsection{Two-Loop Renormalization and RG Invariants}

It might be interesting to see if this $\kappa=0$ Einstein manifold is a fixed point of the $\kappa$ flow to all orders. We will calculate the two-loop correction using the well known formula in terms of the Riemann tensor $R^{\alpha}_{\,\,\beta\gamma\delta}$ (see e.g. \cite{Ketov}),
\begin{align}
\mu\der{}{\mu}g^{(2)}_{\rho\sigma}(\kappa,\lambda)=\frac{1}{8\pi^2}R_{\rho\alpha\beta\gamma}(\kappa,\lambda)R_{\sigma}^{\,\,\,\alpha\beta\gamma}(\kappa,\lambda).
\label{renormGroup2}
\end{align}

The components of the Riemann tensor may be found similarly to the components of the Ricci tensor above using the methods of Appendix A (in particular making use of \eqref{connectionExplicit},\eqref{riemannAppendix}). Then the the two independent parameters in the metric flow according to the equations,

\begin{align}
&\mu\der{}{\mu}\left(\frac{1}{\lambda^2}\right)=\frac{1}{\pi}(\mathcal{N}-1+\kappa)\frac{N}{\mathcal{N}}+\frac{\lambda^2}{2\pi^2}\left(4\mathcal{N}-6N\frac{N}{\mathcal{N}}(1-\kappa)+(3\mathcal{N}-1)\left(\frac{N}{\mathcal{N}}\right)^2(1-\kappa)^2\right),\\
&\mu\der{}{\mu}\left(\frac{1-\kappa}{\lambda^2}\right)=\frac{1}{\pi}(\mathcal{N}-1)\frac{N}{\mathcal{N}}(1-\kappa)^2+\frac{\lambda^2}{2\pi^2}(\mathcal{N}-1)\left(\frac{N}{\mathcal{N}}\right)^2(1-\kappa)^3.
\end{align}

For $\kappa=1$, these equations reduce to the known 2-loop beta function for the Grassmannian model \cite{BrezinEtAl1980}. For $\mathcal{N}=N$, they reduce to the 2-loop beta function for the fibered $CP(N-1)$ model first found in \cite{MouhannaEtAl1995}.
These two-loop equations were also checked using coordinate methods for the special case of $L=M=2$.

Note that for general Grassmannians ($\mathcal{N}\neq N$) the two-loop term is not the same between the two RG equations if we set $\kappa=0$. This means that counter-intuitively, although $\kappa=0$ is indeed an Einstein manifold, higher loop corrections cause $\kappa$ to run.

Finally, for completeness, let us return to the simpler one-loop case, and generalize the results of \cite{MouhannaEtAl1995} on RG invariants. Apart from the ambiguity of the $\kappa=0$ manifold, the qualitative behavior of the one-loop RG equations is much the same as for the fibered $CP(N-1)$ case treated in \cite{MouhannaEtAl1995}. As before \eqref{Kdef}, there is an invariant $K$ which is constant along different RG trajectories, 
\begin{align}
K= \frac{\kappa^{1-\frac{1}{\mathcal{N}}}}{1-\kappa}\lambda^2.
\end{align}
Using this to eliminate $\lambda^2$ from the RG equation for $\kappa$ \eqref{RGKappaGrass}, we can integrate to find an implicit equation for $\kappa$ as a function of the scale $\mu$ in terms of the hypergeometric function $_2 F_1$,
\begin{align}
K \ln\frac{\mu}{\mu_0}&=A(\kappa(\mu))-A(\kappa(\mu_0)),\label{hypergeom}\\
A(\kappa)&\equiv \frac{\mathcal{N}}{N}\frac{\pi}{\mathcal{N}-1}\kappa^{1-\frac{1}{\mathcal{N}}}\,_2F_1\left(2,1-\frac{1}{\mathcal{N}};2-\frac{1}{\mathcal{N}};\kappa\right).
\end{align}
If we define the IR scale $\Lambda$ such that,
$$A(\kappa(\mu_0))=K \ln\frac{\mu_0}{\Lambda},$$
then inserting in \eqref{hypergeom} we see that,
$$A(\kappa(\mu))=K\ln\frac{\mu}{\Lambda},$$
so the definition of $\Lambda$ does not depend on the particular scale $\mu$. 
\begin{align}
\Lambda = \mu \exp\left[-\frac{A(\kappa(\mu))}{K}\right].
\end{align}
Since $A(\kappa)>0$, our one-loop theory is clearly invalid for $\mu<\Lambda$, and as usual $\Lambda$ has the physical interpretation as the order of magnitude at which non-perturbative effects become large, and we would expect it to be of the same order of magnitude as the mass gap.

\section{Discussion and Conclusion  }
\label{dac}

In this paper we studied a continuous class of sigma models on a space which is a $U(1)$ fibration over Grassmannian models, including the special case of complex projective space $CP(N-1)$. We examined these models by pulling the metric back to the Lie group $SU(N)$ in which the left invariance of the metric became clear. The combined restrictions of left-invariance and gauge invariance limited us to just two parameters, $\lambda$ which is the usual coupling constant describing the overall size of the manifold, and $\kappa$ which describes the size of the $U(1)$ fibers. As generically occurs in asymptotically free theories, after quantization the parameter $\lambda$ is transmuted to a dimensionful scale $\Lambda$. The parameter $\kappa$ is also replaced by a RG invariant $K$, which we have shown has an interpretation in the fibered $CP(N-1)$ model as the anomalous dimension of the complex unit-vector $n$ field.

One might try to extend the construction in this paper even further by fibering Grassmannian manifolds by other subgroups of $SU(N)$ besides $U(1)$. The Grassmannian has a natural $SU(L)\times SU(M)\times U(1)$ gauge symmetry, and we can fiber the model by breaking any combination of these commuting subgroups. That is, we could use a Lagrangian which is an extension of \eqref{lagrangianGrassJ}, introducing parameters $k_L, k_M, k_\phi$ for the $SU(L), SU(M), U(1)$ directions respectively, 
\begin{align}\Lagr=\frac{1}{2\lambda^2}\left(\sum_{a\in \text{K\"ahler}}(J^a)^2+k_L\sum_{a\in SU(L)}(J^a)^2+k_M\sum_{a\in SU(M)}(J^a)^2+k_\phi(J^\phi)^2\right).\nonumber\end{align}
The explicitly broken gauge symmetry no longer constrains our set of parameters, but one would expect the global symmetry is sufficient. The RG equations for this model can be found via a straightforward calculation with the formula \eqref{mainFormulaBody}. As a simplified example for the sake of discussion we will present the RG equations for a less general model which is a $SU(2)$ fibered $Gr(2,4)$ model, with $k_L=k_\phi=0$,
$$\mu\der{}{\mu}\lambda^2=-\frac{\lambda^4}{2\pi}\left(8 -\frac{3}{2}k_M\right)$$
$$\mu\der{}{\mu}k_M=\frac{\lambda^2}{2\pi}\left(2+\frac{5}{2}k_M^2-8k_M\right).$$
Unlike the $U(1)$ fibered model, there is a problem here. When we set $k_M=0$ in the Lagrangian we get the Lagrangian for the ordinary Grassmannian model. This is analogous to setting $\kappa=1$ in the $U(1)$ fibered model. But in this case where there are non-Abelian fibers, $k_M=0$ is not a fixed point of the $k_M$ RG equation. There are indeed other fixed points for $k_M$ which do correspond to Einstein manifolds, but $k_M$ appears to flow away from zero, so it is not clear in what sense this can be considered an extension of the Grassmannian model.

The problem is the $k_M$ independent term in the RG equation, which comes from non-vanishing structure coefficients of the $SU(M)$ subgroup. Such a term would be there even if we were considering a trivial geometry in which there is no interaction between the $SU(M)$ fibers and the base manifold, and it can be understood as due to the curvature of the $SU(M)$ Lie group itself.

The non-trivial problem with non-Abelian fibers is that as the parameter $k_M$ goes to zero, the curvature of the fibers themselves diverges, which corresponds to large values of the associated coupling constant $\lambda^2/k_M$. So we do not expect that the RG equations found above are valid near $k_M=0$. Thus the non-Abelian fibered model in the regime near the ordinary Grassmannian model can not be investigated using the perturbative methods of this paper.

However the $U(1)$ fibered model considered here can be investigated perturbatively near the $\kappa=1$ regime close to the Grassmannian. The situation is similar to the difference between the free $O(2)$ sigma model and the higher $O(N)$ models. In the special case of complex projective space $CP(N-1)$, we are able to use the validity of perturbation theory in this regime to find an expression for the two-point correlation function \eqref{powerLaw}.

\section*{Acknowledgments}
The authors are grateful to Andrey Losev, Alexander Voronov, Paul Wiegmann, and Arkady Vainshtein for useful discussions. This work is supported in part by DOE grant DE-SC0011842.

\section*{Appendices}

\begin{appendices}
	\section{Geometry from structure coefficients}
	\label{Geometry}

This appendix will discuss the geometry of manifolds that can be considered to be the orbit of an action by a Lie group, such as those considered in this paper. The group action induces a pull-back map that lets us consider the metric on the Lie group itself. Typically this metric will be degenerate in the sense that there are directions in the Lie group space that have vanishing norm. But in the case that the metric is left-invariant, such as those considered in this paper, we will be able to use the algebraic properties of the Lie group to determine the geometry of the manifold we are interested in.

In section \ref{mfb} we will discuss how the curvature of the degenerate metric on the Lie group determines the curvature on the original manifold. The main result we will need is equation \eqref{riemannTensorEquality} which states that the components of the Riemann tensors of the two spaces in the horizontal directions are equal. This will allow us in Sect. \ref{crt} to use the left-invariance property of the metric on the group space to give a much simpler formula for the Ricci tensor, in an approach similar to that of Milnor \cite{Milnor}. To illustrate the abstractions in section \ref{mfb} in a concrete setting, in section \ref{appendix so3} various properties of $S^{N-1}$ are calculating using the group $SO(N)$.

\renewcommand{\theequation}{A.\arabic{equation}}
\setcounter{equation}{0}
\subsection{Gauge invariant metrics on fiber bundles}
\label{mfb}
\subsubsection{Push-forwards and pull-backs}
\label{mm}

For the moment let us abstract slightly. We have a fiber bundle $\mathcal{E}=\SU(N)$ which maps to the base space $\mathcal{M}=S^{2N-1}$. The projection map $\pi:\mathcal{E}\rightarrow\mathcal{M}$ is given concretely by \eqref{Udef}, which says for a $U\in\mathcal{E}$,
\begin{align}
\pi(U)=U n_0.
\end{align}
In the other direction, we may choose a section $\sigma:\mathcal{M}\rightarrow\mathcal{E}$, that maps each element of the base space to a particular element in the fiber. Of course $\sigma$ is required to be compatible with the projection in the sense that the composition $\pi\circ\sigma$ is just the identity. Concretely $\sigma$ encodes our choice of unitary matrix $U$ for each unit vector $n$.
\begin{align}
\sigma(n(x))=U(x).
\end{align}
We can use these maps to push-forward and pull-back objects living on $\mathcal{E}$ and $\mathcal{M}$. In particular, the metrics on the two spaces are also related by these maps. Given a metric $g$ on $\mathcal{M}$, we can use $\pi^\star$ to pull it back to a degenerate metric $\bar{g}$ on $\mathcal{E}$,
\begin{align}
\bar{g}\equiv\pi^\star g\label{metricPullBack}
\end{align}
This choice of metric is manifestly \emph{gauge invariant} in the sense that $\bar{g}$ nowhere depends on the choice of section $\sigma$.

We can also recover $g$ from $\bar{g}$ by using $\sigma$. This follows since for any curve $\gamma$ associated with a vector in $T\mathcal{M}$, we have by definition of the section, $\pi\circ\sigma\circ\gamma=\gamma.$
This implies $\pi_\star\circ \sigma_\star$ is the identity map $i_{T\mathcal{M}}$ on $T\mathcal{M}$,
\begin{align}
\pi_\star\circ \sigma_\star=i_{T\mathcal{M}}.\label{pushForwardIdentity}
\end{align}
Thus, from the definition of $\bar{g}$ we can recover $g$ by pulling back with any section $\sigma$,
\begin{align}
\sigma^\star\bar{g}= g.
\end{align}

So, our approach will be to consider $\sigma$ as a map locally embedding $\mathcal{M}$ as a submanifold of $\mathcal{E}$. The intrinsic metric $g$ is induced naturally as a pull-back of $\bar{g}$. This intrinsic metric doesn't depend on the details of the embedding map $\sigma$, which is another formulation of the gauge invariance property of $\bar{g}$.

Considering $\mathcal{M}$ as a submanifold in this way, we will derive a close analogue of the Gauss equation which relates the curvature of a submanifold $\mathcal{M}$ to the curvature of the ambient space $\mathcal{E}$ (see for instance \cite{doCarmo}).



\subsubsection{Connections}
\label{conn}

To define the curvature tensor we need to introduce a connection. In the case of $\mathcal{M}$, the Riemannian connection $\nabla$ is uniquely determined by the metric $g$ as usual. But since $\bar{g}$ is degenerate, metric compatibility and vanishing torsion are not enough to specify the connection $\bar{\nabla}$ on  $\mathcal{E}$ uniquely, as we will see explicitly later. For now, the lack of uniqueness will not be a problem, and $\bar{\nabla}$ is any metric compatible and torsion-free connection. Defining $\partial_X$ as the directional derivative in the $X$ direction, we can write
\begin{align}
\partial_X \bar{g}(Y,Z)&=\bar{g}(\bar{\nabla}_X Y,Z)+\bar{g}(Y,\bar{\nabla}_X Z)\,,\label{metricCompat}\\[2mm]
[X,Y]&=\bar{\nabla}_X Y-\bar{\nabla}_Y X\,.&\label{torsionFree}
\end{align}

Now at each point of the submanifold $\sigma(\mathcal{M})$ we can decompose the tangent space $T\mathcal{E}$ into a \emph{parallel} space tangent to the submanifold $\sigma_\star (T\mathcal{M})$ and a \emph{vertical} space $(T\mathcal{E})^\perp$ consisting of all those vectors $\eta \in (T\mathcal{E})^\perp$ such that $\pi_\star \eta=0$,
$$T\mathcal{E}= \sigma_\star (T\mathcal{M})\oplus (T\mathcal{E})^\perp\,.$$
In passing, note that the distinction between the term \emph{parallel} and the previously used term \emph{horizontal} is that the parallel directions depend on the choice of map $\sigma$ and the horizontal directions depended on our choice of left invariant basis $\tau_a$. We can choose $\sigma$ so that the these two notions are identical at any given point of the submanifold, but by the Frobenius theorem it is impossible for them to be the same at every point of $\mathcal{M}$ since the Lie brackets of the horizontal vector fields are not closed.

The key relation between the connections is that the parallel component of the covariant derivative of two parallel vector fields on $\mathcal{E}$ is just the Riemannian covariant derivative on $\mathcal{M}$,
\begin{align}
\pi_\star(\bar{\nabla}_{{\bar{X}}}{\bar Y})={\nabla}_X Y.\label{connectionconnection}
\end{align}
Here $\bar{X}, \bar{Y}$ are local extensions in $T\mathcal{E}$ which agree with $\sigma_\star X, \sigma_\star Y$ on the submanifold. It is straightforward to show different extensions agree when restricted to the submanifold, so the right hand side is well defined.

To prove this relation, first note that $\pi_\star(\bar{\nabla}_{{\bar X}}{\bar Y})$ at the very least indeed defines some valid connection ${\nabla}^\prime$ on $\mathcal{M}$.
\begin{align*}
{\nabla}^\prime_{{X}}{Y}&\equiv\pi_\star(\bar{\nabla}_{{\bar X}}{\bar Y})\end{align*}
By the previous comment ${\nabla}^\prime$ is well-defined acting on vectors in $T\mathcal{M}$, and from the linearity of the push-forward, it is linear and obeys the Leibniz product rule under multiplication by a scalar.

The fact that indeed ${\nabla}^\prime=\nabla$ follows from the uniqueness of the Riemannian metric. We can prove metric compatibility for ${\nabla}^\prime$ by using the metric compatibility \eqref{metricCompat} for $\bar{\nabla}$ and the definition of $\bar{g}$ as a pull-back \eqref{metricPullBack}. We can prove the vanishing of torsion from \eqref{torsionFree}, since the push-forward of the Lie bracket of vector fields is equal to the Lie bracket of the push-forward. This last statement relies on the fact that all vector fields involved are parallel to the submanifold.

Equation \eqref{connectionconnection} which we just proved means that $\bar{\nabla}_{{\bar X}}{\bar Y}$ can be decomposed into
parallel and vertical components,
\begin{align}
\bar{\nabla}_{{\bar X}}{\bar Y}=\sigma_\star{\nabla}_X Y+\eta,\label{connectionPerp}
\end{align}
where $\eta\in(T\mathcal{E})^\perp$.

Metric compatibility implies a curious feature about these vectors $\eta$. By the pull-back definition of the metric \eqref{metricPullBack}, and the definition of the $(T\mathcal{E})^\perp$ as the kernel of $\pi_\star$, we see that for any vector $X$,
$$\bar{g}(X,\eta)=0.$$
Conversely a vector $\eta$ satisfying this property must be in $(T\mathcal{E})^\perp$ since $\pi_\star \eta$ has zero norm in the non-degenerate metric $g$. If we take the covariant derivative of the expression above,
$$\bar{g}(\bar{\nabla}_Y X,\eta)+\bar{g}(X,\bar{\nabla}_Y \eta)=\bar{g}(X,\bar{\nabla}_Y \eta)=0,$$
we see that for all vectors $Y$
\begin{align}
\bar{\nabla}_Y \,\eta\in(T\mathcal{E})^\perp.\label{perpVectorDer}
\end{align}
So the space of vertical vector fields is closed under taking covariant derivatives in any direction.

\subsubsection{Riemann and Ricci tensors}
\label{rrt}

The Riemann tensor is defined as a map on vectors,
\begin{align}
R(X,Y)=\nabla_{[X,Y]}-\nabla_X\nabla_Y+\nabla_Y\nabla_X.
\end{align}
As before we will distinguish $R$ on $\mathcal{M}$ and $\bar{R}$ on $\mathcal{E}$ by use of a bar.

The Ricci tensor is defined as a trace over an arbitrary basis $Z_a$. Because the metric $\bar{g}$ is degenerate, rather than using an orthonormal basis, let us simply work with a compatible basis of dual vectors $\hat{Z}^a$,
\begin{align}
\hat{Z}^a(Z_b)=\delta^a_b.
\end{align}
Then the Ricci tensor is defined as
\begin{align}
\text{Ric}(X,Y)\equiv\sum_a \hat{Z}^a\left(R(X,Z_a)Y\right).
\end{align}

The key result used in the calculation of the Ricci tensor is that when $X,Y,Z,W$ are all parallel to the submanifold, the coefficients of the Riemann tensors of the two spaces are equal,
\begin{align}
\hat{\bar W}(\bar{R}(\bar X,\bar Y)\bar Z)=\hat{W}({R}(X,Y)Z). \label{riemannTensorEquality}
\end{align}
To prove this, consider the difference of the two sides of the equation. Identifying vectors and dual vectors on $\mathcal{M}$ with the corresponding parallel objects in $\mathcal{E}$ in the obvious way (i.e. omitting use of $\sigma_\star$ for brevity),
\begin{align*}
\hat{\bar W}(\bar{R}(\bar X,\bar Y)\bar Z-{R}(X,Y)Z)&=\hat{\bar W}(\bar{\nabla}_{[\bar{X},\bar{Y}]}\bar{Z}-\nabla_{[X,Y]}Z-\bar{\nabla}_{\bar{X}}\bar{\nabla}_{\bar{Y}}\bar{Z}+\nabla_X\nabla_Y Z+\dots).
\end{align*}
By \eqref{connectionPerp}, the difference of the first two terms involving the derivative in the Lie bracket direction is some element $\eta$ in the vertical direction. So these terms vanish when acted on by the parallel $\bar{W}$. 

In the third and fourth term, using first \eqref{connectionPerp} then \eqref{perpVectorDer}, \begin{align*}
-\bar{\nabla}_{\bar{X}}\bar{\nabla}_{\bar{Y}}\bar{Z}+\nabla_X\nabla_Y Z&=-\bar{\nabla}_{\bar{X}}(\nabla_Y Z +\eta)+\nabla_X\nabla_Y Z\\[2mm]
&=-(\bar{\nabla}_{\bar{X}}\nabla_Y Z -\nabla_X\nabla_Y Z)+\eta^\prime\,,
\end{align*}
where $\eta^\prime$ is some new vertical vector. Both of these terms are vertical and so vanish when acted on by $\bar{W}$. Similarly the remaining terms vanish the same way and so this proves the equality \eqref{riemannTensorEquality}.

Note finally that we derived this relation by using a map $\sigma$, but the only appearance of $\sigma$ in \eqref{riemannTensorEquality} is in the notion of what it means to be a parallel vector. At each point this equality must be true for any possible notion of parallel. So in the following we will simply consider the vectors $X,Y,Z,W$ to be in the space spanned by the horizontal left invariant vector fields.

\subsection{Calculation of the Ricci tensor}
\label{crt}

Now that we have shown the curvature of the degenerate metric $\bar{g}$ on the fiber bundle directly determines the curvature of $g$ on the base space, we can use the Lie group structure of the fiber bundle to simplify the calculation of the Ricci tensor. The key properties which make this simplification possible are that the metric $\bar{g}$ is left-invariant and that the Lie bracket of the left-invariant vector fields $\tau_a$ are just isomorphic to the commutator of the Lie algebra elements that they correspond to
\begin{align}
[\tau_a,\tau_b]_L\equiv \nabla_{\tau_a}\tau_b-\nabla_{\tau_b}\tau_a=-2\sum_c f_{abc}\tau_c.\label{tauBracket}
\end{align}
We must be a little careful in that the matrix commutator is between the \emph{anti}-Hermitian matrices which have absorbed a factor of $i$. For this reason we use a subscript L to indicate that this should be considered the Lie derivative of $\tau$ as vector fields, which is almost but not quite the same as the the commutator of $\tau$ considered to be the Hermitian matrices discussed earlier. The normalization of the structure coefficients $f$ is chosen to agree with the standard where the Lie algebra basis elements involve an extra factor of $1/2$ compared to the normalization in \eqref{traceOrthogonal}.

Following Milnor \cite{Milnor}, we will use these structure coefficients of the Lie algebra to determine the connection coefficients of the manifold.

To begin recall that metric is diagonal in our choice of basis,
\begin{align}
\bar{g}(\tau_a,\tau_b)=\frac{1}{\lambda^2}C_a \delta_{ab}.\label{metricLI}
\end{align}
In particular, from \eqref{lagrangianJ} we have that $C_a=0$ for the vertical directions, $C_a=1$ for the  K\"ahler directions, and
\begin{align}
C_{N^2-1}=(1-\kappa)\frac{2(N-1)}{N},\label{metricPhase}
\end{align}
for the phase direction.

Since the metric is constant in this basis, by metric compatibility
\begin{align*}
\bar{g}(\bar{\nabla}_{\tau_a} \tau_b,\tau_c)+\bar{g}(\tau_b,\bar{\nabla}_{\tau_a} \tau_c)=0.
\end{align*}
Then by repeatedly using the vanishing torsion condition \eqref{torsionFree}, we can derive a relation in terms of Lie brackets, which we then can write in terms of structure coefficients \eqref{tauBracket} and metric components \eqref{metricLI},
\begin{align}
\bar{g}(\bar{\nabla}_{\tau_a} \tau_b,\tau_c)&=\frac{1}{2}\left[\,\,\bar{g}([\tau_a,\tau_b]_L,\tau_c)-\bar{g}([\tau_b,\tau_c]_L,\tau_a)+\bar{g}([\tau_c,\tau_a]_L,\tau_b)\,\,\right]\non
&=-\frac{1}{\lambda^2}f_{abc}\left(C_c-C_a+C_b\right).
\end{align}
Note that we used the fact that the structure coefficients are completely antisymmetric in our choice of Lie algebra basis. Also no summation convention over repeated Lie algebra indices is implied in this section.

The connection must respect this equation and also the torsion-free condition, which implies
\begin{align}
\bar{\nabla}_{\tau_a} \tau_b=-\sum_c f_{abc}\left(1+\frac{C_b-C_a}{C_c}\chi_c\right)\tau_c + \eta_{ab}.\label{connectionExplicit}
\end{align}
Here $\chi_c$ is an indicator function which is $1$ on horizontal indices and $0$ on vertical indices. $\eta_{ab}$ is an arbitrary set of vectors belonging to the vertical subspace which are symmetric under permutation of $a, b$. This is the non-uniqueness of the connection for degenerate metrics mentioned previously. By the theorem on the Riemann tensor \eqref{riemannTensorEquality}, the choice of $\eta$ will not affect our calculation of the Riemann tensor on the base space, and so in the following we will simply take $\eta=0$.  

Now the components of the Riemann tensor are given by
\begin{align}
\hat{\tau}^d(R(\tau_a,\tau_b)\tau_c)=\bar\tau^d\left(\left[\nabla_{[\tau_a,\tau_b]_L}-\nabla_{\tau_a}\nabla_{\tau_b}+\nabla_{\tau_b}\nabla_{\tau_a}\right]\tau_c\right),\label{riemannAppendix}
\end{align}
and then by taking the trace in $d$ and $b$ and using \eqref{tauBracket} and \eqref{connectionExplicit}, we can find the diagonal components of the Ricci tensor,
\begin{align}
R_{aa}	&=\sum_{b,c}f_{abc}^2\left(1+\frac{C_b-C_a}{C_{{c}}}\chi_c+3\frac{C_a-C_c}{C_{{b}}}\chi_b-\frac{C_b-C_a}{C_{{c}}}\frac{C_a-C_c}{C_{{b}}}\chi_b\chi_c\right).\label{mainFormula}
\end{align}
So this formula only depends on information about the group through $f_{abc}$ and the constant components of the metric $C_a$. There is no explicit dependence on target space position unlike the coordinate method in Appendix \ref{appendix coordinates}.

To actually calculate with this, let us review the relevant structure coefficients for $\SU(N)$. First of all, the standard basis for the Lie algebra contains elements of the Cartan subalgebra which we will denote $\tau_{k'}$
\begin{align}
\tau_{k'}\equiv\sqrt\frac{2}{k(k-1)}\text{diag}(1,1,\dots,1,-(k-1),0,\dots,0) 
\end{align} 
Here $k$ can range from $2$ to $N-1$, i.e. we are specifically not including the phase element $\tau_{N^2-1}$.

These Cartan subalgebra elements appear in structure coefficients between \emph{paired}  K\"ahler elements as in \eqref{tauLastRowColumn}. By ``paired'' we mean both $\tau_{M+2j-1}$ and $\tau_{M+2j}$ with the same $j$,
\beq
f_{M+2j-1, \,\,M+2j,\,\, k'}= \left\{\begin{array}{lc}
	0\,,  &  \quad k<j  \\[2mm]
	-\sqrt{\frac{k-1}{2k}}\,, & \quad k=j \\[2mm]
	\frac{1}{\sqrt{2k(k-1)}}\,, & \quad k>j
\end{array}\right.
\label{strucCoeff}
\eeq
\beq
f_{M+2j-1, \,\,M+2j,\,\, N^2-1}=\sqrt{\frac{N}{2(N-1)}}\,.
\label{strucCoeffLast}
\eeq
Also there are structure coefficients between \emph{unpaired}  K\"ahler elements. For any unpaired $i$ and $j$ there is exactly one $k$ from the vertical subalgebra with nonvanishing structure coefficient, 
\begin{align}
f_{M+i, \,\,M+j,\,\, k}=\pm \frac{1}{2}, \label{strucCoeffUnpaired}
\end{align}
where the particular sign will not be relevant in our calculation.

Up to permutation, these are the only nonvanishing structure coefficients involving the $2N-1$ horizontal elements. In the formula for the Ricci tensor \eqref{mainFormula}, this has the consequence that for any horizontal $a$, any term involving a vertical $b$ must vanish since then both $a$ and $c$ are  K\"ahler elements and $C_a=C_c=1$.

Now we expect that $R_{aa}$ is the same for each  K\"ahler element $a$, just as is the case for the metric $g_{aa}$. If this were not the case there would need to be extra parameters in the Lagrangian, which we have already argued conflicts with combined gauge and left invariance.

So let us calculate for one of the last pair of  K\"ahler elements, $a=N^2-1-2$. Then we have four contributions to the sum \eqref{mainFormula}, namely, 
\begin{enumerate}[(i)]
	\item There are $2(N-2)$ unpaired generators $b$, involving structure coefficient \eqref{strucCoeffUnpaired}. Each term contributes $+1$, so it contributes in total,
	$2(N-2)\,;$ \item For $b=N^2-1$, $c$ must be the paired generator as in \eqref{strucCoeffLast}. This contributes
	$(1-\kappa)\,;$
	\item For $b$ paired, $c=N^2-1$, which contributes
	$\frac{2N}{N-1}-3(1-\kappa)\,;$
	\item\label{lastTerm} Finally, for this particular choice of $a$ the only nonvanishing term involving a member of the vertical Cartan subalgebra is $c=(N-1)'$, which involves \eqref{strucCoeff}. And so this contributes $\frac{2(N-2)}{N-1}.$ 
\end{enumerate}
So, summing these four terms we can find the  K\"ahler components of the Ricci tensor,
\begin{align}
R_{aa}=2(N-1+\kappa).\label{Rproj}
\end{align}

On a sidenote, it is only the last term (\ref{lastTerm}) which might change for a different value of  K\"ahler element $a$. Even if the reader is not persuaded by the argument given that $R_{aa}$ must be the same for any  K\"ahler index $a$, it is straightforward to check that it indeed is the same by making use of the identity
$$\sum_{i=2}^{k-1}\frac{1}{i(i-1)}=\frac{(k-2)}{k-1}.$$

Now the only remaining component of the Ricci tensor is the phase component, which can be found by a similar but slightly shorter calculation with \eqref{mainFormula},
\begin{align}
R_{N^2-1,\,\,N^2-1}=\frac{2(N-1)}{N}2(N-1)(1-\kappa)^2.\label{Rphase}
\end{align}
These components of course agree with the Ricci tensor calculated straightforwardly via explicit coordinates in Appendix \ref{appendix coordinates}.
The Ricci scalar is obtained by convoluting (\ref{Rproj}) and (\ref{Rphase}) with the inverse metric,
\beq
R= 2\lambda^2\,(N-1) \left(\kappa +2N -1\right).
\label{tue1}
\eeq

		 \subsection{Geometry of \boldmath{$S^{N-1}$} via \boldmath{$SO(N)$}}
		 \label{appendix so3}

		 As a more familiar example of the methods and notation used in this Appendix, consider $SO(3)$ as fiber bundle $\mathcal{E}$ with base space $\mathcal{M}$ being the unit sphere $S^2$. The projection map $\pi$ sends orthogonal matrices $O\in SO(3)$ to a unit vector $n\in S^2$, by acting on the reference unit vector $n_0=(0,0,1)$,
		 $$\pi(O)=O n_0=n.$$
		 A general element orthogonal matrix $O$ that maps to $n$ has the form,
		 \begin{align}O=	\left(\begin{array}{ccc}
		 e_{(1)} & e_{(2)} & n
		 \end{array}\right)\label{Ogeneral}\end{align}
		 where $e_{(1)},e_{(2)}, n$ are orthonormal column vectors.

		 The left invariant vector field which is associated to a Lie algebra element $\tau$ is given at each point $O$ by differentiation along the path $\bar{\gamma}$ parametrized by $\theta$, $$\bar{\gamma}(\theta)=O \exp(\theta\tau).$$ In the present case, the factors $\exp(\theta\tau)$ associated to the standard anti-Hermitian generators $\tau_x,\tau_y,\tau_z$ are just the rotation matrices about the $x, y, z$ axes respectively.  So using the explicit rotation matrices it is easy to find the push-forward of the left invariant vector fields at a given point $O$,
		 \begin{align}
		 \pi_*(\tau_x)&=-e_{(2)}\non
		 \pi_*(\tau_y)&=+e_{(1)}\non
		 \pi_*(\tau_z)&=0.
		 \end{align}
		 
		 $e_{(1)},e_{(2)}$ are orthonormal in the unit sphere metric inherited from $R^3$. So $\tau_x, \tau_y$ are also orthonormal vectors in the pull-back metric $\bar{g}$. Since the pull-back metric does not depend on the point $O$, this metric is left invariant.
		 
		 To map vectors in the other direction from $S^2$ to $SO(3)$ we need choose a section $\sigma$ that maps each point on the sphere to an orthogonal matrix. For instance, in standard polar coordinates $\theta, \phi$ on $S^2$ we might make the choice,
		 \begin{align}
		 \sigma(\theta, \phi)=\left(\begin{array}{ccc}
		 \cos\phi\cos\theta & -\sin\phi & \cos\phi\sin\theta\\
		 \sin\phi\cos\theta & \cos\phi & \sin\phi\sin\theta\\
		 -\sin\theta & 0 & \cos\theta
		 \end{array}\right)
		 \end{align}
		 
		 The push-forward of the coordinate vectors $\partial_\theta, \partial_\phi$ are just given by the derivative along the path $\sigma(\theta,\phi)$ as the respective coordinate is varied. This can be expressed in a basis of left-invariant vector fields by finding the anti-Hermitian matrix $J=\sigma^{-1}\partial \sigma$ associated with the path.
		 
		 \begin{align}
		 \sigma_* \partial_\theta &= \sigma^{-1} \partial_\theta \sigma = \tau_y\non
		 \sigma_* \partial_\phi &= \sigma^{-1} \partial_\phi \sigma = -\sin\theta \tau_x +\cos\theta\tau_z 
		 \end{align}
		 
		 Note that by \eqref{pushForwardIdentity} acting on the coordinate vectors by $\pi_* \circ \sigma_*$ should be the identity. In this case we indeed recover the coordinate vectors from the columns $e_{(1)}, e_{(2)}$ of $\sigma$.
		 
		 Now let's consider how the covariant derivatives on the two spaces are related. The covariant derivative $\bar{\nabla}$ on $SO(3)$ is determined by \eqref{connectionExplicit},
		 $$\bar{\nabla}_{\tau_a} \tau_b=-\sum_c f_{abc}\left(1+\frac{C_b-C_a}{C_c}\chi_c\right)\tau_c.$$
		 In this case the structure coefficients are just the Levi-Civita symbol (with a normalization chosen to agree with \eqref{tauBracket})
		 $$f_{abc}=-\frac{1}{2}\epsilon_{abc},$$
		 and the diagonal components of the metric are,
		 $$C_x=C_y=1,\qquad C_z=0.$$
		 So explicitly,
		 \begin{align}\bar{\nabla}_{\tau_x} \tau_z=\bar{\nabla}_{\tau_y} \tau_z=0\non
		 \bar{\nabla}_{\tau_z} \tau_x=+\tau_y\non \bar{\nabla}_{\tau_z} \tau_y=-\tau_x\non
		 \bar{\nabla}_{\tau_x} \tau_y=-\bar{\nabla}_{\tau_y} \tau_x=\frac{1}{2}\tau_z\end{align}
		 
		 Now from \eqref{connectionconnection}, the push-forward $\pi_*$ of this covariant derivative is equal to the covariant derivative on the unit sphere. As a curiosity, note we can explicitly find the Christoffel connection coefficients this way. For instance, using the push-forwards $\sigma_*$ of the coordinate vectors, and freely using the fact that $\pi_*(\tau_z)=0$,
		 \begin{align*}
		 \nabla_\phi \partial_\theta &= \pi_*\left(-\sin\theta \bar{\nabla}_{\tau_x}\tau_y +\cos\theta\bar{\nabla}_{\tau_z}\tau_y\right)\\
		 &= \pi_*\left(\cot\theta(-\sin\theta \tau_x+\cos\theta \tau_z) \right)\\
		 &=\cot\theta \,\partial_\phi.
		 \end{align*}
		 From this we see $\Gamma^{\phi}_{\phi\theta}=\cot\theta$ and $\Gamma^{\theta}_{\phi\theta}=0$. The other connection coefficients can be calculated similarly.
		 
		 The calculation of the Ricci curvature is even simpler than this since it does not rely on all the machinery of choosing a section $\sigma$ and finding the push-forward maps. Using the formula \eqref{mainFormula}, with $a$ one of the two horizontal directions,
		 \begin{align*}
		 R_{aa}	&=\sum_{b,c}f_{abc}^2\left(1+\frac{C_b-C_a}{C_{{c}}}\chi_c+3\frac{C_a-C_c}{C_{{b}}}\chi_b\right)\\
		 &=\frac{1}{4}\left(1+\frac{0-1}{1}\right)+\frac{1}{4}\left(1+3\frac{1-0}{1}\right)=1
		 \end{align*}
		 The Ricci scalar then traces over the two diagonal directions, so it is simply $R=2$.
		 
		 This calculation can be trivially extended to $SO(N)$ acting on $S^{N-1}$, since the structure coefficients are still just proportional to the Levi-Civita symbol. Now members of the horizontal directions are the $N-1$ antisymmetric matrices which are only non-zero in the last row and column. Given a horizontal direction $a$ in the formula above, $b$ can be any of the other $N-2$ directions, so the calculation is generalized to $R_{aa}=N-2$, and the trace over $N-1$ elements leads to
		 $${\cal R}=(N-1)(N-2),$$
		 which is of course the correct Ricci scalar for the unit sphere $S^{N-1}$.
		 
	\section{Explicit coordinate methods}
	\label{Coordinate}
	\subsection{A special case of \boldmath{$N=2$}}
	\label{appendix A}
	 \renewcommand{\theequation}{B.\arabic{equation}}
	 \setcounter{equation}{0}
		 
		 The $N=2$ case is special in that $SU(2)\sim S^3$ so there is no vertical subgroup, and no gauge fixing is necessary to define the unitary matrix representation. In this case it is worth studying a parametrization of $U$ which is routinely used in the $\SU(2)\times \SU(2)/\SU(2)$ chiral models for pions, namely,
		 \beqn
		 U &=& \left(1 + i \, \frac{\pi^a\tau^a}{2} \right) \left(1 - i \,\frac{\pi^a\tau^a}{2} \right)^{-1}\,,
		 \nonumber\\[2mm]
		 U^\dagger &=& \left(1 + i \, \frac{\pi^a\tau^a}{2} \right)^{-1} \left(1 - i \frac{\pi^a\tau^a}{2} \right)\,,
		 \label{w85}
		 \eeqn
		 where the summation over $a=1,2,3$ is implied. Then
		 \beqn
		 J_\mu &=& U^\dagger \partial_\mu U =  {i}\, \left(1 + i \,\frac{\pi^a\tau^a}{2} \right)^{-1}\left(\partial_\mu\pi^b\tau^b\right) \left(1 - i \,\frac{\pi^c\tau^c}{2}\right)^{-1}
		 \nonumber\\[4mm]
		 &=&  {i}\,  \Big(1 + \frac{\pi^2}{4}\Big)^{-2}\,
		 \left(1 - i \,\frac{\pi^a\tau^a}{2} \right)\left(\partial_\mu\pi^b\tau^b\right) \left(1 + i \,\frac{\pi^c\tau^c}{2}\right),
		 \label{w86}
		 \eeqn
		 where
		 \beq
		 \pi^2 = \pi^a\pi^a\,.
		 \label{w87}
		 \eeq
		 The individual components $J^a_\mu$ can be found by taking traces with Pauli matrices $\tau^a$ as in \eqref{cl2}. The result is,
		 \begin{align}
		 J_\mu^a=\left(1 + \frac{\pi^2}{4}\right)^{-2}\left[\left(1 - \frac{\pi^2}{4}\right)\partial_\mu \pi^a +\left(\pi \times \partial_\mu \pi \right)^a +\frac{1}{2}\left(\pi \cdot \partial_\mu \pi\right)\pi^a\right].
		 \end{align}
		 
		 These components appear in the Lagrangian based on \eqref{cl1}
		 \begin{align}
		 \Lagr = \frac{1}{2\lambda^2}\left(\sum_a \left(J^a_\mu\right)^2 -\kappa\left(J^3_\mu\right)^2\right)
		 \end{align}
		 The first term which does not depend on $\kappa$ is just the Lagrangian for the PCM,
		 \begin{align}
		 \frac{1}{2\lambda^2}\left(\sum_a \left(J^a_\mu\right)^2\right)=\frac{\left(\partial_\mu \pi^a\right)^2}{2\lambda^2\left(1 + \frac{\pi^2}{4}\right)^2}.
		 \end{align}
		 The deformation term breaks the symmetry between $\pi^1,\pi^2$ and $\pi^3$. Let us introduce the notation
		 \begin{align}
\phi \equiv \pi^1 + i \pi^2 \qquad \sigma\equiv \pi^3
		 \end{align}
		 Then to quartic order, the deformation terms in the Lagrangian are
		 \begin{eqnarray}
		-\frac{\kappa}{2\lambda^2}\left(J_\mu^3\right)^2 
		&=& 
		-\frac{\kappa}{2\lambda^2}\Big[\partial^\mu\sigma\partial_\mu \sigma-  i\partial^\mu \sigma\left(\phi^\dagger\partial_\mu \phi - \phi \partial_\mu \phi^\dagger\right) -\frac{1}{2}\sigma^2(\partial\sigma)^2\nonumber\\[2mm]
		 &+&
		 \frac{1}{2}|\phi|^2|\partial\phi|^2-\frac{1}{4}\left(\phi^{\dagger 2}(\partial_\mu \phi)^2+\phi^{2}(\partial_\mu\phi^\dagger)^2\right)\nonumber\\[2mm]
		&-&\frac{3}{2}|\phi|^2(\partial\sigma)^2+\frac{1}{2}\sigma\partial^\mu \sigma \left(\phi^\dagger\partial_\mu \phi + \phi \partial_\mu \phi^\dagger\right)+{O}(\pi^5,\pi^6)\Big].
		 \end{eqnarray}
%
%
%
%

		 \subsection{Extension of Fubini-Study coordinates}
		 \label{appendix coordinates}

		 An alternate method to finding the renormalization group equations is simply to choose an unconstrained coordinate system and try to calculate the Ricci tensor directly from connection coefficients.
		 
		 As mentioned earlier, we will use a simple extension of the Fubini-Study coordinates on $CP(N-1)$. The real and imaginary components of the Fubini-Study coordinates are given by $$z^i=x^i+i y^i=\frac{n^i}{n^0},$$
		 where $x$ and $y$ are real.
		 
		 To these $2(N-1)$ coordinates we also add the extra coordinate $\phi$ parametrizing the overall phase of $n$. For convenience we will also define the quantity
		 \begin{align}\chi\equiv 1+|z|^2.\end{align}
		 Then by transforming the Lagrangian \eqref{lagrangianN} to these coordinates we can read off the components of the metric,
		 \begin{align}
		 g_{\phi\phi}&=1-\kappa\,,\non
		 g_{\phi x_i}&=-(1-\kappa)\chi^{-1}y_i\,,\non
		 g_{\phi y_i}&=+(1-\kappa)\chi^{-1}x_i\,,\non
		 g_{x_i x_j}&=\chi^{-1}\delta_{ij}-\chi^{-2}\left(x_i x_j+\kappa y_i y_j\right)\,,\non
		 g_{y_i y_j}&=\chi^{-1}\delta_{ij}-\chi^{-2}\left(y_i y_j+\kappa x_i x_j\right)\,,\non
		 g_{x_i y_j}&=-\chi^{-2}\left(x_i y_j - \kappa y_i x_j\right)\,.
		 \end{align}
		 If we stare at this long enough we can guess and check the components of the inverse metric,
		 \begin{align}
		 g^{\phi\phi}&=\frac{1+(1-\kappa)|z|^2}{1-\kappa}\,,\non
		 g^{\phi x^i}&=\chi y^i\,,\non
		 g^{\phi y^i}&=-\chi x^i\,,\non
		 g^{x^i y^j}&=\chi (x^iy^j-y^ix^j)\,,\non
		 g^{x^i x^j}=g^{y^iy^j}&=\chi(\delta^{ij}+x^ix^j+y^iy^j)\,.
		 \end{align}
		 
		 Now it is straightforward to calculate the connection coefficients,
		 \begin{align}
		 \Gamma^\phi_{\phi\phi}&=\Gamma^{x_i}_{\phi\phi}=\Gamma^{y_i}_{\phi\phi}=\Gamma^{x_j}_{\phi x_i}=	\Gamma^{y_j}_{\phi y_i}=0\,,\non[2mm]
		 \Gamma^\phi_{\phi x_i}&=-(1-\kappa)\frac{x_i}{\chi},\qquad\Gamma^\phi_{\phi y_i}=-(1-\kappa)\frac{y_i}{\chi}\,,\non[2mm]
		 \Gamma^{y_j}_{\phi x_i}&=+(1-\kappa)\delta_{ij},\qquad\Gamma^{x_j}_{\phi y_i}=-(1-\kappa)\delta_{ij}\non\Gamma^\phi_{ x_i y_j}&=
		 \frac{\kappa}{\chi^2}(x_i x_j-y_i y_j)\,,\non[2mm]
		 \Gamma^\phi_{ x_i x_j}&=-	\Gamma^\phi_{ y_i y_j}=-\frac{\kappa}{\chi^2}(x_i y_j+y_i x_j)\,,\non[2mm]
		 \Gamma^{x_k}_{ x_i x_j}&=-\chi^{-1}(\delta_{ik}x_j+\delta_{jk}x_i),\qquad\Gamma^{y_k}_{ x_i x_j}=\kappa\chi^{-1}(\delta_{ik}y_j+\delta_{jk}y_i)\,,\non[2mm]
		 \Gamma^{y_k}_{ y_i y_j}&=-\chi^{-1}(\delta_{ik}y_j+\delta_{jk}y_i),\qquad\Gamma^{x_k}_{y_i y_j}=\kappa\chi^{-1}(\delta_{ik}x_j+\delta_{jk}x_i)\,,\non[2mm]
		 \Gamma^{x_k}_{ x_i y_j}&=-\chi^{-1}(\delta_{ik}y_j+\kappa\delta_{jk}y_i),\qquad\Gamma^{y_k}_{ x_i y_j}=-\chi^{-1}(\kappa\delta_{ik}x_j+\delta_{jk}x_i)\,.	\end{align}
		 
		 From this point a short route to the RG equations is to calculate not the full Ricci tensor but only the tensor at the point $z=0$, and only calculate the $R_{\phi\phi}$ and $R_{x_1x_1}$ components. From the general argument that only the $\lambda$ and $\kappa$ parameters should renormalize, this shorter calculation gives us all the information about the full Ricci tensor.
		 
		 At $z=0$ the metric becomes diagonal in this coordinate system,
		 \beq
		 \left(g\right)_{0}=\lambda^{-2}\text{diag}(1,1,\dots,1,1-\kappa),
		 \eeq
		 where the final component is the one associated to the $\phi$ coordinate.
		 
		 Calculating from the connection coefficients at $z=0$, the Ricci tensor components are
		 \beqn
		 R_{x_1 x_1}(0)&=&2(N-1+\kappa)\\[2mm]
		 R_{\phi\phi}(0)&=&2(N-1)(1-\kappa)^2.
		 \eeqn
		 These are indeed equal to the components in the left invariant basis \eqref{Rproj} and \eqref{Rphase} respectively, keeping in mind the proportionality between $\tau_{N^2-1}$ and the $\phi$ direction in \eqref{phiCoord}. So, using the form of the metric at $z=0$, we of course calculate the same RG equations as \eqref{lambdaFlow} and \eqref{kappaFlow}.	
	
\end{appendices}

    \end{document}